\documentclass[times,twocolumn,final]{elsarticle}

\usepackage{lipsum}
\makeatletter
\def\ps@pprintTitle{%
	\let\@oddhead\@empty
	\let\@evenhead\@empty
	\def\@oddfoot{}%
	\let\@evenfoot\@oddfoot}
\makeatother

\usepackage{framed,multirow,hhline}
\usepackage{amsmath,amssymb,amsfonts}
\usepackage[margin=0.65in]{geometry}

\usepackage{natbib}
\usepackage{hyperref}
\hypersetup{
	colorlinks=true,
	citecolor={Cyan},
	linkcolor={Red},
	urlcolor={Cyan}}

\begin{document}	
	
	\begin{frontmatter}
		
		\title{Automatic Segmentation and Location Learning of Neonatal Cerebral Ventricles in 3D Ultrasound Data Combining CNN and CPPN}
		
		\author[1]{Matthieu Martin}
		\corref{cor1}
		\cortext[cor1]{Corresponding author: matthieu.martin@creatis.insa-lyon.fr}
		
		\author[1]{Bruno Sciolla}
		
		\author[1]{Michael Sdika}
		
		\author[2]{Philippe Quetin}
		
		\author[1]{Philippe Delachartre}
		
		\address[1]{Univ Lyon, INSA‐Lyon, Université Claude Bernard Lyon 1, UJM-Saint Etienne, CNRS, Inserm, CREATIS UMR 5220, U1206, F‐69621, LYON, France}
		\address[2]{CH Avignon, AVIGNON, France}
		
		\begin{abstract}
			Preterm neonates are highly likely to suffer from ventriculomegaly, a dilation of the Cerebral Ventricular System (CVS). This condition can develop into life-threatening hydrocephalus and is correlated with future neuro-developmental impairments. Consequently, it must be detected and monitored by physicians. In clinical routing, manual 2D measurements are performed on 2D ultrasound (US) images to estimate the CVS volume but this practice is imprecise due to the unavailability of 3D information. A way to tackle this problem would be to develop automatic CVS segmentation algorithms for 3D US data. In this paper, we investigate the potential of 2D and 3D Convolutional Neural Networks (CNN) to solve this complex task and propose to use Compositional Pattern Producing Network (CPPN) to enable Fully Convolutional Networks (FCN) to learn CVS location. Our database was composed of 25 3D US volumes collected on 21 preterm nenonates at the age of $35.8 \pm 1.6$ gestational weeks. We found that the CPPN enables to encode CVS location, which increases the accuracy of the CNNs when they have few layers. Accuracy of the 2D and 3D FCNs reached intraobserver variability (IOV) in the case of dilated ventricles with Dice of $0.893 \pm 0.008$ and $0.886 \pm 0.004$ respectively (IOV = $0.898 \pm 0.008$) and with volume errors of $0.45 \pm 0.42$ cm$^3$ and $0.36 \pm 0.24$ cm$^3$ respectively (IOV = $0.41 \pm 0.05$ cm$^3$). 3D FCNs were more accurate than 2D FCNs in the case of normal ventricles with Dice of $0.797 \pm 0.041$ against $0.776 \pm 0.038$ (IOV = $0.816 \pm 0.009$) and volume errors of $0.35 \pm 0.29$ cm$^3$ against $0.35 \pm 0.24$ cm$^3$ (IOV = $0.2 \pm 0.11$ cm$^3$). The best segmentation time of volumes of size $320 \times 320 \times 320$ was obtained by a 2D FCN in $3.5 \pm 0.2$ s.
		\end{abstract}
		
	\end{frontmatter}
	
	\section{Introduction}
	\subsection{Motivations}
	Millions of babies are born preterm every year in the world. These neonates, particularly those with a very low birth weight ($<$ 1500 g), are highly likely to suffer from ventriculomegaly (VM) which corresponds to a dilation of the cerebral ventricular system (CVS). This pathology has several etiologies \cite{leviton1996ventriculomegaly} whose most commons are intraventricular hemorrhage (IVH) \cite{burstein1979intraventricular} and white matter damages \cite{pappas2018neurodevelopmental}. 

	In grade 3 and 4 IVH \cite{papile1978incidence}, CVS dilation is directly caused by the hemorrhage which prevents the CerebroSpinal Fluid (CSF) produced inside the CVS from flowing out normally. This case is particularly critical because VM can develop into hydrocephalus, in which case the increased intracranial pressure can cause convulsions, brain damages and lead to death. When VM is caused by white matter damages, the ventricles are dilated to compensate for the loss of brain white matter volume and there is no risk of hydrocephalus. In both cases, studies carried out on US scans \citep{ancel2006cerebral, fox2014relationship, larroque2003white} and on MRI scans \cite{melhem2000periventricular, maunu2011ventricular} have shown a correlation between VM and future neuro-developmental impairments such as cerebral palsy or delayed cognitive performance.\\
	
	To mitigate the short-term risks and the long-term impairments for these patients, VM must be detected and the ventricular volume (VV) monitored. Even though it is easy to detect VM when the CVS is very dilated, quantitative longitudinal measurement of its volume is required to evaluate the severity of the dilation and its resorption. Based on these parameters, clinicians can prognosticate for future impairments and decide to perform a surgical intervention \cite{mazzola2014pediatric}.	Recently, \cite{leijser2018posthemorrhagic} has shown that an early surgical intervention was more appropriate than a late intervention to limit neurodevelopmental impairments. Therefore, early detection of VM and precise and regular monitoring of VV are of great clinical interest.
	
	To image preterm neonates’ CVS, 2D transfontanellar sonography is the most common clinical routine examination. Compared to the other imaging modalities, it is low cost, highly available and low-risk for babies. Clinicians estimate the CVS volume by performing various manual measurements, on these 2D images, such as the ventricular index or the anterior horn’s widths \cite{brouwer2012new, davies2000reference}. However this methodology is imprecise: the estimated volume is not correctly correlated with direct volume measurements \cite{kishimoto2016preterm} performed on 3D US data and the intra/inter-observer dependency can make the evaluation of the CSF volume evolution difficult. A good correlation was recently shown between measurements performed on MRI scans and on 3D US data \cite{boucher2018computer} and measurements made on in 3D US data were reported to be more accurate and faster \cite{kishimoto2016quantitative} than those performed on 2D US data. This suggest that 3D US exams have a great potential for this application. In that case, automatic and fast CVS segmentation algorithms need to be developed.

	\subsection{Related works}
	 Several methods have been proposed to achieve preterm neonates' CVS segmentation in MRI data. Automatic pixel-to-pixel segmentation methods were developped by \cite{moeskops2016automatic} and \cite{liu2017combining}. \cite{moeskops2016automatic} proposed a CNN which performs segmentation based on features from three independant branches with different patch size as input and \cite{liu2017combining} combined spatial and non-spatial dictionnary learning to achieve CVS segmentation. An active contour based method was also proposed by \cite{qiu20153d}. It used a multiphase geodesic level-set which utilizes a spacial prior obtained from multi-atlas registration. Unfortunately, that imaging modality is not suitable to monitor all preterm neonates because of its low accessibility and its inconvenience for babies. Conversely, US scans are available and convenient for these patients. But US images are hard to analyze because of their low contrast, the presence of speckle and the realization of the acquisition in an unknown coordinate system. Several methods have been proposed to achieve preterm neonate’s CVS segmentation in US data.

	In 2D US data, a semi-automatic segmentation algorithm which uses a method based on an active contour which incorporates a prior on the location was proposed by \cite{sciolla2016segmentation}. Automatic segmentation was achieved by \cite{tabrizi2018automatic} applying a combination of fuzzy c-mean, phase congruency and active contour-based method and by \cite{yasarla2019learning} using a 2D-CNN-based method which combines segmentation blocks and confidence blocks. 
	
	In 3D US data, its segmentation was achieved semi-automatically by \cite{qiu2015user}. This algorithm used manually labeled points as initialization and an intensity-based surface-evolution method to perform CVS segmentation. Automatic segmentation methods were proposed by \cite{boucher2018dilatation}, \cite{qiu2017automatic} and \cite{martin2018automatic}. \cite{boucher2018dilatation} used a multi-altas registration algorithm to initialize segmentation which was then performed using a deformable mesh model. The authors reported Dice and MAD in \cite{boucher2018dilatation} on their own dataset containing 12 infants whose age ranged from $2$ to $8.5$ months. Respective values of $0.708 \pm 0.036$ and $0.88 \pm 0.2$ mm were obtained. The segmentation time was not reported. \cite{qiu2017automatic} also initialize CVS segmentation utilizing a multi-atlas registration and performed segmentation with a multi-phase geodesic level-set method based on phase congruency maps. Their personal database contained $14$ patients with IVH grade I to III whose gestational age ranged from $37$ to $42$ weeks. The authors reported a Dice, a MAD and a segmentation time of $0.767 \pm 0.062$, $1.0 \pm 0.3$ mm and $54$ min per volume, respectively. Finally, \cite{martin2018automatic} achieved CVS segmentation with a 2D U-net on a subset of the US database (15 patients) used in this article. The latter has the advantage of being very fast (only a few seconds compared to 54 minutes).\\ 
	
	The 3D US data segmentation algorithms described in this article are based on CNNs. The later have recently outperformed the state-of-the-art methods dedicated to image segmentation \cite{badrinarayanan2015segnet, long2015fully}. In particular, the use of CNNs for medical image semantic segmentation enables to reach outstanding levels of accuracy \cite{milletari2016v, ronneberger2015u} in a clinical time. Deep learning methods enable to learn relevant features directly from data. It is interesting in difficult problems, such as CVS segmentation, where the anatomical structure has a complex shape (Fig.\ref{fig:data}.d) and wide range of intensity (Fig.\ref{fig:data}.a).
	
	Common CNN architectures, such as V-net and U-net, do not encode information about absolute location since convolution operations are equivariant to translation. In laymen terms, a traditional CNN attempts to classify the tissue manually, as if a physician did it using a small picture patch. It has no clue whether it comes from inside the brain, the cortex prefrontal area or the cerebellum area. Nevertheless, the automation of the conventional methods used in 3D ultrasound to segment CVS \cite{boucher2018dilatation, qiu2017automatic} and in MRI to segment several brain structures \cite{srhoj2013automatic, makropoulos2014automatic} rely on the use of atlases. The latter allow to introduce an a priori on the position of the cerebral structures. Therefore, it seems important to estimate the benefit of a spatial a priori in the context of CVS segmentation by CNNs. In that case, the use of atlases did not seem relevant to us because only MRI atlases are avaible for this application. Their use could result in imprecise and difficult registrations because the data comes from different imaging modalities and because of the huge anatomical variability at that age. Moreover, atlas registration would no longer allow for segmentations in a clinical time of a few seconds. To introduce a spatial a priori in CNNs, \cite{ghafoorian2017location, de2015deep, ganaye2018towards} have proposed to concatenate spatial features to the input of the classification layer. Nevertheless, these approaches rely on handcrafted spatial features which does not enable the CNNs to learn abstract spatial a priori.
	
	\subsection{Contributions}
	 In this paper, we investigate the potential of 2D and 3D CNNs for CVS segmentation in 3D US data. We evaluate the benefits of using 2D FCN instead of 2D CNN which achieve pixel-to-pixel segmentation. In addition, we compared 2D and 3D FCNs. Finally, we propose to use Compositional Pattern Producing Network (CPPN) \cite{stanley2007compositional} to enable the FCNs to learn CVS location. 
	
	CPPNs are specific types of networks outputting shapes from input such as coordinates. Combined with CNNs low-level features, they can provide information about the features’ location within the head. This approach has the advantage to learn the CVS-location directly from our dataset and enable the networks to be end-to-end trained. The accuracy of the networks was evaluated using two usual metrics (Dice and Mean Absolute Distance (MAD) and two medical metrics (absolute and relative volume difference). 
	
	 In the following sections, we show that the addition of the CPPN enables the learning of a pattern dictionary that improves the accuracy of the FCNs when they have few layers. We also show that 2D FCN are more accurate and faster than 2D CNN which achieve pixel-to-pixel segmentation. In addition, we show that 2D and 3D FCN can segment dilated CVS at IOV level and that 3D FCNs are more accurate than 2D FCNs in the case of normal CVS. Both architectures enable CVS segmentation in a clinical time of a few seconds. The main contributions of this paper are:
	
	\begin{enumerate}
		\item U-net and V-net learning spatial information end-to-end with the use of a CPPN, resulting in improved accuracies when they have few layers.
		\item The comparison of 2D and 3D CNNs for CVS segmentation in 3D US data.
		\item The first automatic CVS segmentation in 3D US data, using 3D CNN, performed in a clinical time.
	\end{enumerate}
	
	\section{Materials and Methods}
	\subsection{US data description}
	\subsubsection{Data acquisition}
	\begin{figure}[!b]
		\centering
		\includegraphics[scale=.25]{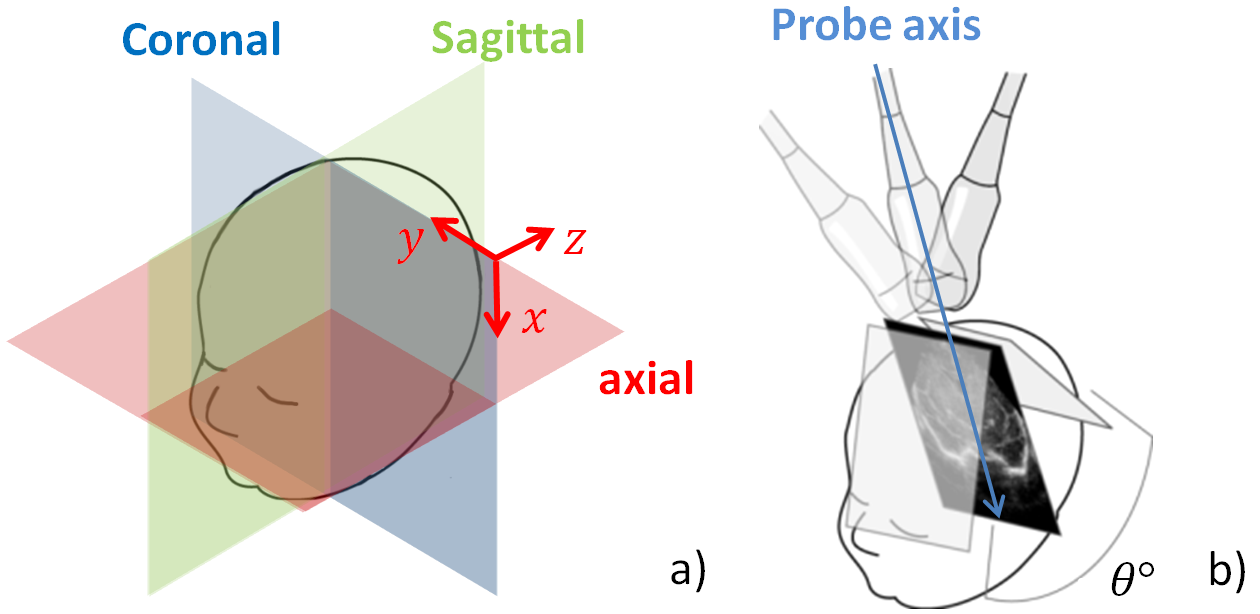}
		\caption{a): anatomical reference planes, b): transfontanellar 2D US acquisition: manual sweep of angle $\theta$.}
		\label{fig:acquisition}
	\end{figure}

	For this study, 2D transfontanellar sonography acquisitions were performed on 21 preterm neonates at the neonatology center of the hospital of Avignon in France. These acquisitions corresponded to angular manual sweeps (Fig.\ref{fig:acquisition}.b) performed through the anterior fontanel with an Acuson Siemens 9L4 multi-D matrix transducer. A total of 25 acquisitions were obtained, each of them contained in average $249 \pm 69$ images of size $567 \times 763$, their spatial resolution was $0.15$ mm/pixel. The mean age of the infants at acquisition and at birth were respectively $35.8 \pm 1.6$ and $31.9 \pm 2.9$ gestational weeks. 
		
	\subsubsection{Data pre-processing}
	
	\begin{figure}[!t]
		\centering
		\includegraphics[scale=.4]{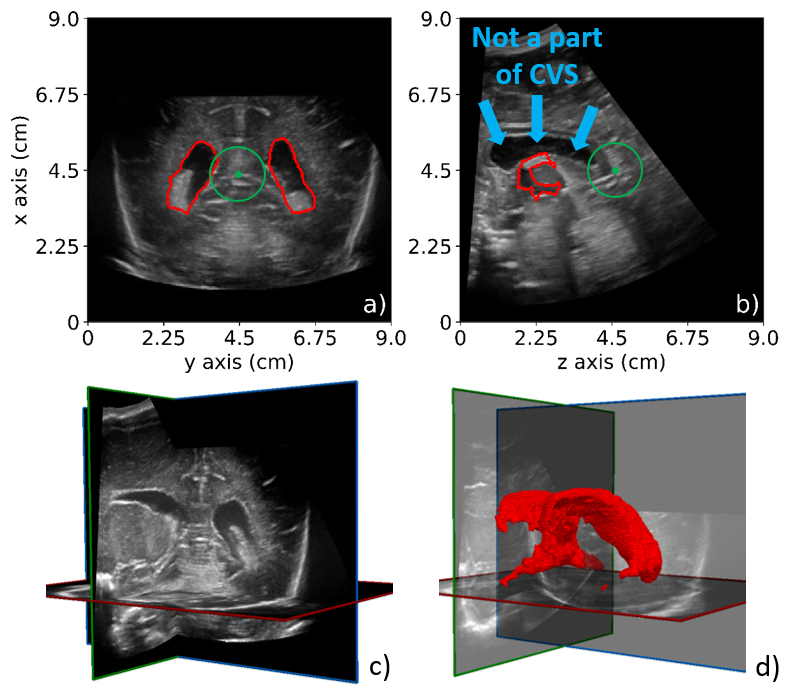}
		\caption{a), b): coronal and sagittal view of the corpus callosum splenium (green dot) and the CVS borders (red lines), the Cavum Septi Pellucidi (blue arrow) is also visible in b). c): 3D US reconstructed volume, d): 3D CVS manual segmentation.}
		\label{fig:data}
	\end{figure}

	3D-reconstructed US volumes were obtained from the 2D acquisitions with the reconstruction algorithm we described in \cite{martin2018automatic}. After 3D reconstruction (Fig.\ref{fig:data}.c), the data was manually centered on the corpus callosum splenium (green points in Fig.\ref{fig:data}.a.b) and rotated to the standard MRI coordinate system with axes $x$, $y$ and $z$ (Fig.\ref{fig:acquisition}.a). Then, the volumes were scaled to the half of their original size using bicubic interpolation and cropped to $320 \times 320 \times 320$ voxels. This permitted defining a common coordinate system for all volumes and then computing identically normalized coordinate maps used as input for the CPPN. Finally, each image (respectively volume) were standardized to constitute the database used for 2D networks (respectively 3D networhs).
		
	\subsubsection{CVS description and data annotation}
\begin{figure}[!h]
		\centering
		\includegraphics[scale=.55]{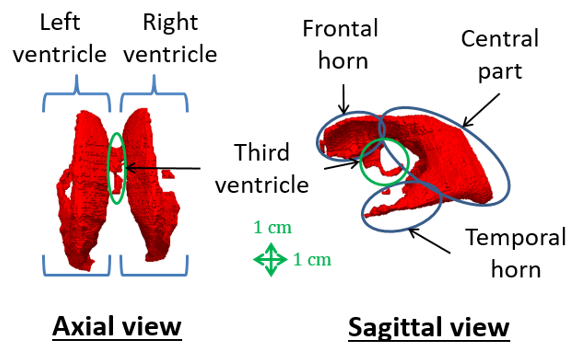}
		\caption{Anatomical description of the CVS in axial and sagittal view.}
		\label{fig:CVS}
	\end{figure}

	CVS segmentation is part of the difficult segmentation problems: it deals with complex spatial structures with variable aspects in terms of image contrast. Fig.\ref{fig:data}.a shows that the CVS is composed of a hyperechoic part and a hypoechoic part. The last one must not be confused with the Cavum Septi Pellucidi (CSP) visible in Fig.\ref{fig:data}.b which is also hypoechoic. The complex anatomy of the CVS, which includes very thin parts, is described in Fig.\ref{fig:CVS}.
	
	Manual CVS segmentations were performed on the US data acquisitions by a physician who checked all the reconstructed annotated volumes (Fig.\ref{fig:data}.d) for anatomical and spatial consistency. These reference segmentations were used to measure each CVS volume, its mean value and the number of normal and dilated CVS composing the dataset are given in the last row of Table.\ref{tab:dataset}.
	
	\subsubsection{Dataset creation}
	The volumes were divided into a training set, a validation set and a test set which contained 13, 5 and 7 volumes respectively. All volumes corresponding to the same patient were placed in the training set. The remaining volumes were randomly distributed so that the proportion of dilated and normal CVS was close in each set. Their characteristics are summarized in Table.\ref{tab:dataset}.
	
	\begin{table}[!h] 	
		\caption{Repartition of the dilated and normal CVS between the sets} 
		\centering 
		\begin{tabular}{ccccc} 
			\hline 
			\multirow{2}{*}{Set} & \multicolumn{2}{c}{Normal CVS} & \multicolumn{2}{c}{Dilated CVS} \\ 
			& Number & VV (cm$^3$) & Number & VV (cm$^3$) \\ 
			\hline 
			Training & 9 & 2.7 $\pm$ 0.8 & 4 & 9.4 $\pm$ 0.6 \\ 
			Validation & 3 & 3.1 $\pm$ 0.9 & 2 & 8.4 $\pm$ 2.5 \\ 
			Test & 5 & 2.5 $\pm$ 0.5 & 2 & 9.4 $\pm$ 1.1 \\ 
			All volumes & 17 & 2.7 $\pm$ 0.8 & 8 & 9.1 $\pm$ 1.5 \\ 
			\hline 
		\end{tabular} 
		\label{tab:dataset} 
	\end{table}

	\subsubsection{Four-fold cross validation dataset}
	To prove that the previous definition of the dataset did not usefully bias our results, a four-fold cross validation dataset was defined. In that case, only one volume per patient was retained, the validation set was not modified and the remaining patients were separated into four sets. In each of these sets, the patients were randomly drawn in order to have 1 patient with dilated CVS and 3 patients with normal CVS. This distribution is summarized in Table.\ref{tab:dataset_cross_val}.
	
	\begin{table}[!h] 
		\caption{Repartition of the dilated and normal CVS in the case of the four-fold cross validation dataset} 
		\centering 
		\begin{tabular}{ccccc} 
			\hline 
			\multirow{2}{*}{Set} & \multicolumn{2}{c}{Normal CVS} & \multicolumn{2}{c}{Dilated CVS} \\ 
			& Number & VV (cm$^3$) & Number & VV (cm$^3$) \\ 
			\hline 
			Fold 1 & 3 & 2.5 $\pm$ 0.5 & 1 & 8.69 \\ 
			Fold 2 & 3 & 2.3 $\pm$ 0.4 & 1 & 10.46 \\ 
			Fold 3 & 3 & 3.1 $\pm$ 0.7 & 1 & 9.5 \\ 
			Fold 4 & 3 & 2.0 $\pm$ 0.3 & 1 & 8.33 \\ 
			\hline 
		\end{tabular} 
		\label{tab:dataset_cross_val} 
	\end{table}

	\subsection{MRI data description}
	To make our results comparable, we sought an open-access database that contains 3D manual annotations of the premature child's CVS. Unfortunately, to our knowledge, there is no US database that meets these criteria. The only MRI database meeting these criteria (\cite{ivsgum2015evaluation}) contained few available annotated volumes. We therefore used an MRI database, from the OASIS project (\cite{marcus2007open}) containing a reasonable number of volumes with manual annotations of the CVS of the young adult.
	
	\subsubsection{Data description}
	We used data from the MICCAI challenge on multi-atlas labelling \cite{landman2019miccai}, this database contains 35 annotated volumes divided into 15 training volumes and 20 test volumes. From the 15 training volumes, we used 4 volumes for validation. These images were acquired with a Siemens Vision 1.5T MRI in the sagittal plan at a resolution of $1 \times 1 \times 1.25$ mm per voxel and were resized to a resolution of $ 1 \times 1 \times 1 $ mm. The images were originaly segmented into $134$ classes in coronal plan. For the need of our study, we only retained and fusioned the classes corresponding to CVS.
	
	\subsubsection{Data pre-processing}
	All volumes were re-centered, oriented similarly to US volumes and cropped to the size $320 \times 192 \times 256$. 
	These steps enabled the creation of a common coordinate system for all volumes similar to the one defined for the US database. Finally, each image (respectively volume) was standardized to constitute the database used for 2D networks (respectively 3D networks)

	\subsection{CNN architecture}
	\subsubsection{Deep learning and CNNs}
	In segmentation problems, deep supervised learning aims at learning a conditional probability distribution $P(Y/X)$ where $X$ refers to an input image and $Y$ to a targeted output label map. This approach enables to learn features directly from data. This is particularly suitable for this kind of problem because the CVS has a complex shape and a wide range of pixel intensities, making it hard to define handcrafted features properly. CNNs aim at learning abstract representations of $X$ that can be used by a final classification layer. For this purpose, convolution operations are used to permit learning small features regardless of their absolute locations in an image. This enables the networks to figure out what the CVS boarders look like as well as its inside aspects. This work aims at studying two points: the benefit of using FCNs (\cite{long2015fully}) compared to CNNs with a fully connected classification layer and, among FCNs, if it was more suitable to use a 2D or 3D architecture.\\ 
	
	\subsubsection{FCN}
\begin{figure}[!t]
		\centering
		\includegraphics[scale=.36]{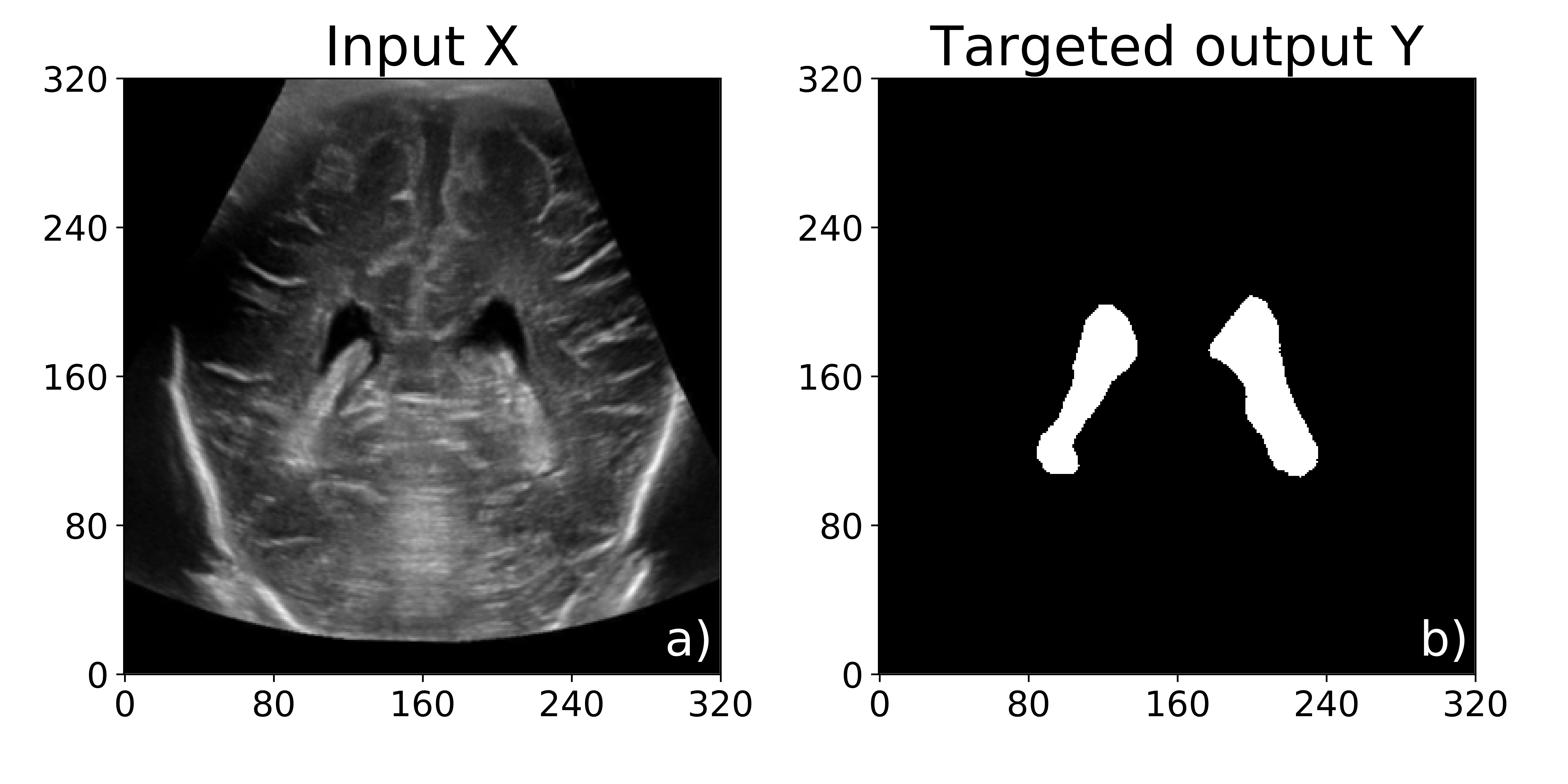}
		\caption{a) Network input X: 2D or 3D US image in coronal view, b) targeted	output Y: 2D or 3D binary image.}
		\label{fig:in_out_data}
	\end{figure}

\begin{figure*}[!t]
		\centering
		\includegraphics[scale=.47]{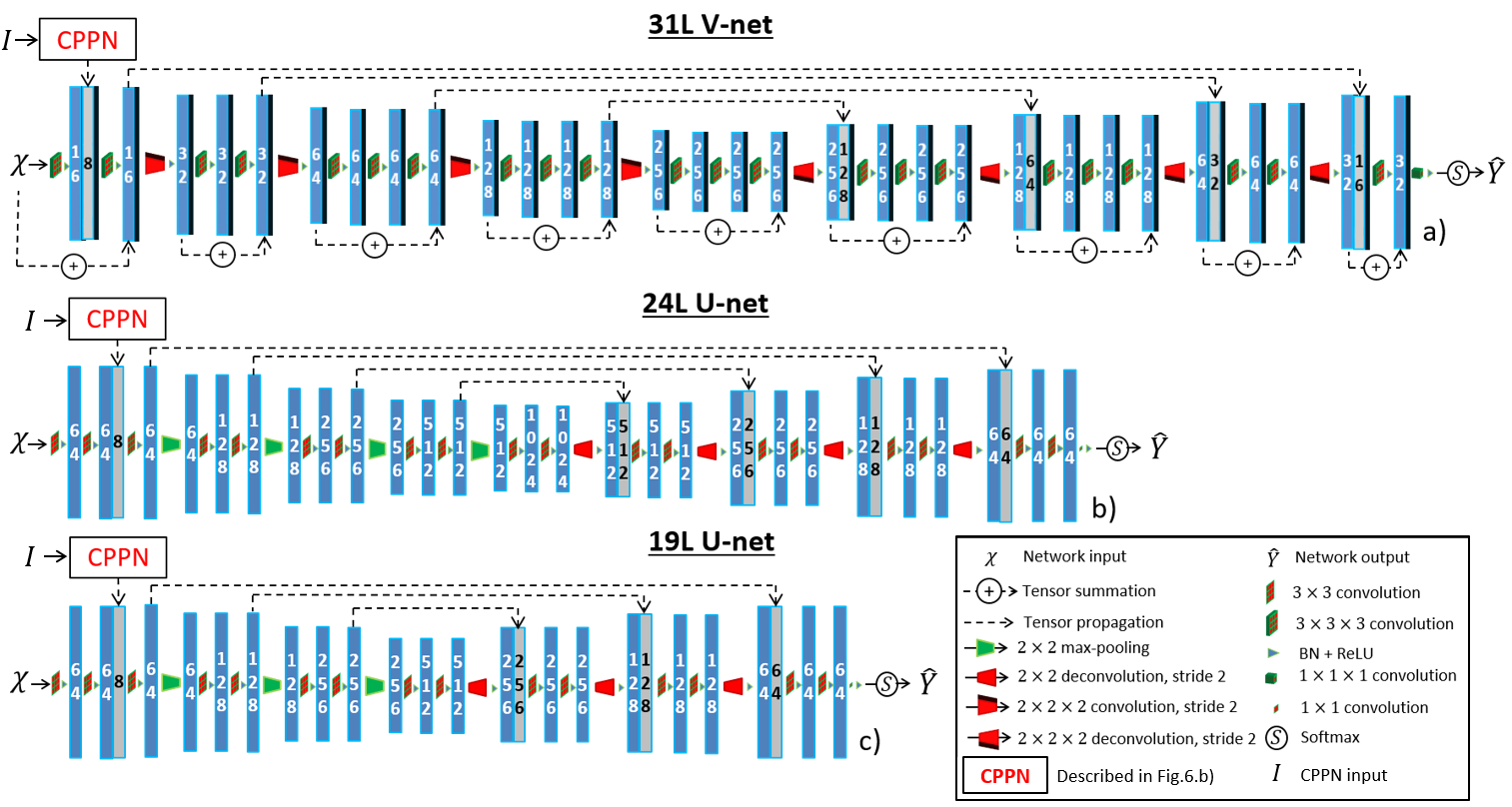}
		\caption{a), b): the originals 24-layer U-net and 31-layer V-net CNNs with CPPN, c): 17-layer U-net with CPPN. The tensors and the concatenated tensors are represented by blue and gray rectangles respectively. Their number of channels is written inside the rectangles.}
		\label{fig:networks}
	\end{figure*}

	FCN are interesting architectures because they enable the full segmentation of an image in one inference. A pixel-wise classification of millions of voxels is then possible in a few seconds only. Among them, we compared V-net and U-net (Fig.\ref{fig:networks}.a and Fig.\ref{fig:networks}.b) which are respectively 3D and 2D efficient architectures to solve these kinds of problems in the field of medical imaging. The inputs $\mathcal{X}$ given to the networks were standardized coronal gray-level images (Fig.\ref{fig:in_out_data}.a) and the targeted outputs $Y$ were the corresponding binary images (Fig.\ref{fig:in_out_data}.b) with the CVS labeled as 1 and the background as 0. The network output was $\hat{Y}$: the pixel-wise probabilities to belong to the background or the CVS. The estimated segmentation maps $\hat{\hat{Y}}$ were eventually obtained by assigning the label with the highest probability to each voxel. In order to be as fair as possible while comparing U-net and V-net for this problem, convolutions of size $3\times3$ for U-net and $3\times3\times3$ for V-net were used as well as $ReLU$ activation for both. Batch Normalization (BN) was included \cite{ioffe2015batch} in the original architectures because it has been reported to improve network parameters optimization by making the optimization landscape smoother \cite{santurkar2018does}. Dropout and the Softmax function were respectively applied just before and right after the classification layer. \\

	\subsubsection{CNN with fully connected classification layer}
	Architectures with fully connected classification layers were used in image segmentation before FCNs. Although they generally offer good accuracy, image segmentation is generally slow because each inference gives a class to only one pixel. Among these architectures, we used the network proposed by \cite{moeskops2016automatic} which has been used for multi-class segmentation of the brain of premature child in MRI. This network is composed of three independent branches that use patches of different sizes (centered on the same pixel) as input. Each branch has three convolutional layers with differently sized convolution that allow learning features at different scales. Each branch ends with a fully connected layer. Finaly, a final fully connected layer gather the characteristics learned by the three branches to achieve classification. In the case of the MRI database, the size of the input patches was $75 \times 75$, $50 \times 50$ and $25 \times 25$ as defined in \cite{moeskops2016automatic}. In the case of the US dataset, as the pixel resolution was smaller, we used sizes of $127 \times 127$, $101 \times 101$ and $75 \times 75$ and consequently adapted the number of neurons in the fully connected layers. The desired output was the class of the pixel in the center of the patches which was represented by 1 for VL and 0 for background. The convolution sizes defined in \cite{moeskops2016automatic} were used. As for U-net and V-net, we used batch normalization and the Softmax function right after the classification layer but no dropout was applied.

	\subsection{CPPN and integration to the networks}
	\subsubsection{CPPNs}
	CPPNs were introduced by \cite{stanley2007compositional} in the field of evolutionary biology to study phenotype evolution. These particular networks use sum and composition of basic mathematical functions to produce geometric patterns from input such as coordinates. Even though CNNs do not encode absolute location information, relevant features can be detected regardless of their positions in a given image thanks to convolution kernels’ specificities. When it can be easy to human beings to determine whether two different features with similar aspects, -such as skull boundary and plexus choroid boundary-, belong to the CVS or not by acknowledging their locations, CNNs have to work differently. They will use contextual information to discriminate them. Nevertheless, this information can be costly to obtain as it requires more layers or bigger convolution kernels. To that end, we defined a CPPN to generate patterns for the CNN to identify and use as spatial-information patterns. In particular, we want the CNN to be able to use the CPPN to remove obvious false positives given their locations. Basically the CPPN component enables networks to learn their own dictionaries of relevant template shapes.
	
	\subsubsection{CPPN designed and integration to U-net and V-net}
	
	\begin{figure}[!t]
		\centering
		\includegraphics[scale=0.39]{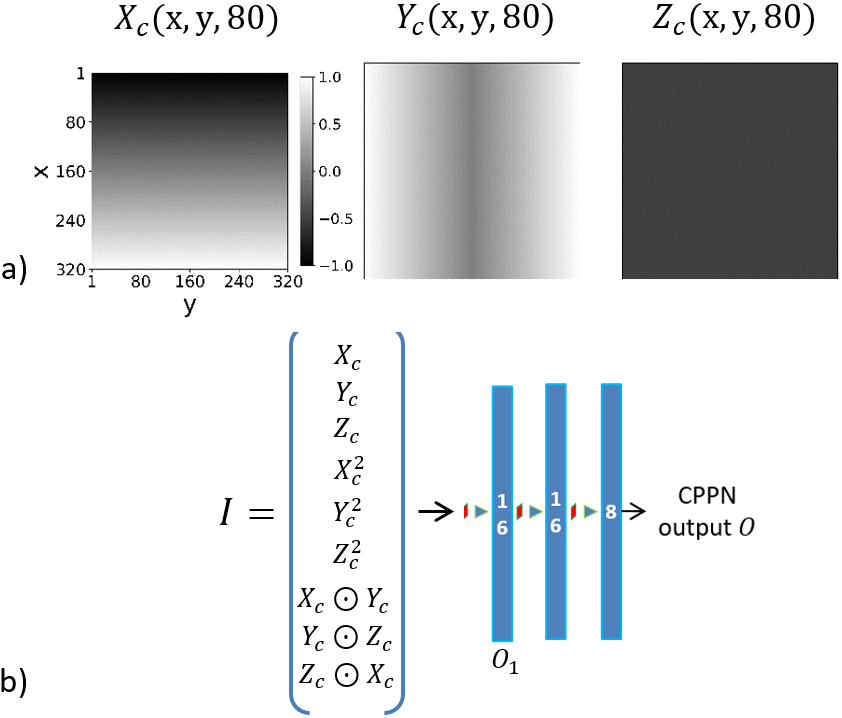}
		\caption{a) Normalized coordinate maps from $X_c$, $Y_c$ and $Z_c$ at a given z, b) CPPN architecture.}
		\label{fig:CPPN_in_CIBM}
	\end{figure}

	\begin{figure*}[!t] 
	\centering
	\includegraphics[scale=0.65]{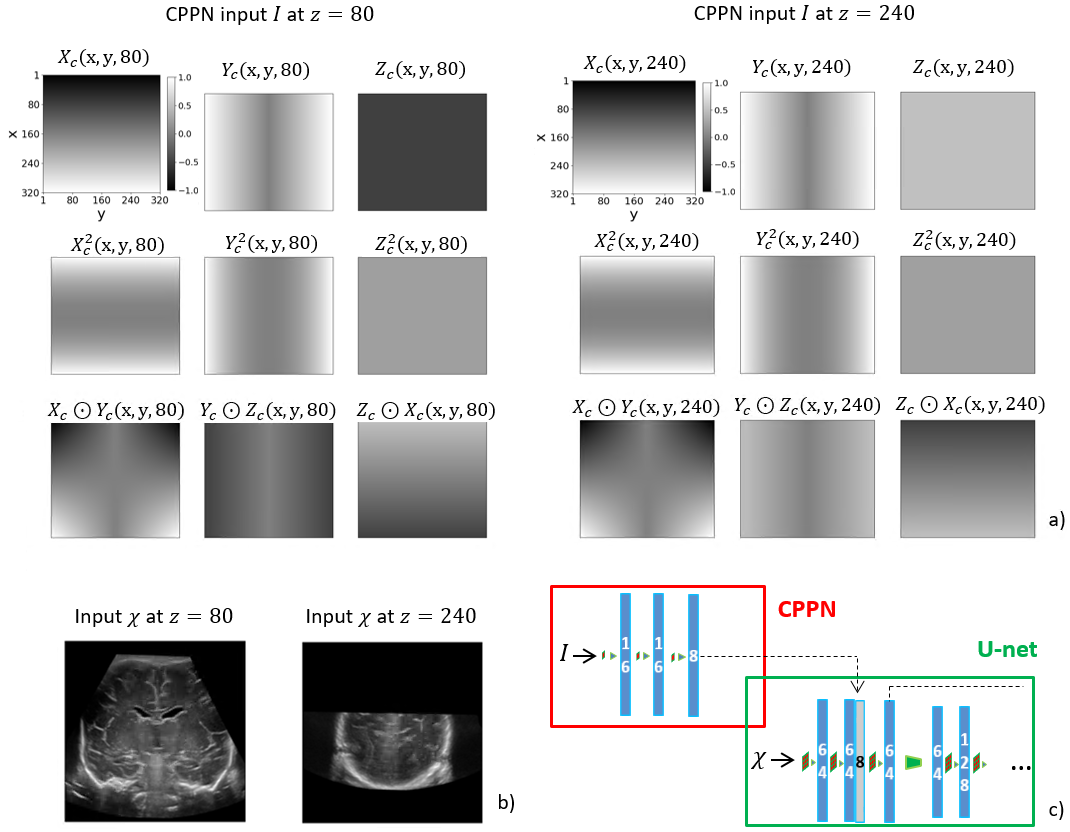}
	\caption{a) Example of CPPN's inputs at $z=80$ and $z=240$ b) Example of U-net's input at $z=80$ and $z=240$, c) integration of the CPPN to U-net.}
	\label{fig:CPPN_in_all_CIBM}
	\end{figure*}
	
	To define our CPPN’s inputs we created normalized-coordinate volumes $X_c$, $Y_c$, $Z_c$. The normalized coordinates $(x_n,y_n,z_n)$ of a point of coordinates $(x,y,z)$ were calculated as follows:
	
	\begin{equation}
	\begin{array}{ccc}
	\vspace{1ex}
	x_n & = & \frac{2x-N_x}{N_x} \\
	\vspace{1ex}
	y_n & = & \vert \frac{2x-N_y}{N_y} \vert\\
	z_n & = & \frac{2x-N_z}{N_z}
	\end{array}
	\label{eqn:norm_coord}
	\end{equation}
	
	Where $N_x$, $N_y$ and $N_z$ correspond to the number of pixel in $x$,$y$ and $z$ axis. Thus, $x_n, z_n \in [-1,1]$ and $y_n \in [0,1]$ which takes into account the symmetry between the right and left hemispheres of the brain and allow for the generation of symmetrical patterns. Finaly, $X_c$, $Y_c$ and $Z_c$ were defined so that for each point of coordinates $(x,y,z)$ :
	
	\begin{equation}
	\begin{array}{ccc}
	(x_n, y_n, z_n) & = & (X_c(x,y,z), Y_c(x,y,z), Z_c(x,y,z))
	\end{array}
	\label{eqn:XYZ}
	\end{equation}
	
	An example of $X_c$, $Y_c$ and $Z_c$ slices for a given $z$ (coronal plan) is shown in Fig.\ref{fig:CPPN_in_CIBM}.a. The CPPN architecture is described in Fig.\ref{fig:CPPN_in_CIBM}.b, the concatenation of the normalized coordinate maps, their squares and their component-wise products (Hadamard : $\odot$) are used as inputs ($I$) to a 3-layer CNN. This CNN’s layers are composed of $1 \times 1 (\times1) $ convolution operations, one BN and then one ReLU activation. In the first layer, the $1 \times 1 (\times1) $ convolutions combine the 9 inputs linearly to generate a quadrics (ellipsoid, hyperboloid, $\hdots$). As an example the i-th output of the first layer $O_{1,i}$ ($O_1$ can be seen in Fig.\ref{fig:CPPN_in_CIBM}.b), before BN and ReLU, is: 
	
	\begin{equation}
	\begin{array}{ccc}
	O_{1,i} & = & w_{i,1} X_c^2 + w_{i,2} Y_c^2 + w_{i,3} Z_c^2 + w_{i,4} X_c \odot Y_c + w_{i,5} X_c \odot Z_c \\
	& & + w_{i,6} Y_c \odot Z_c + w_{i,7} X_c + w_{i,8} Y_c + w_{i,9} Z_c + b_{i}
	\end{array}
	\label{eqn:quad}
	\end{equation}
	
	It is an exact quadratic function of the coordinates with $w_{i,k}$ and $b_i$ being the k-th weights and the bias of the i-th neuron of the first layer respectively. The outputs of the first layer are then combined together by the second and the third layer to create more complex patterns that are non-linear compositions of quadrics. We expect these patterns to be able to help the network figure out where the CVS is located. For instance, some shapes can be identified as regions where the CVS is not likely to be whereas others will be identified as regions where the CVS might be located. However, more complex interactions and logical operations could occur.\\
	
	To enable the U-net and the V-net to be end-to-end trained with the CPPN, the patterns are learnt in parallel of the low-level features before being concatenated with the main network (Fig.\ref{fig:CPPN_in_all_CIBM}.c). Consequently, the extraction of low-level features is kept translation invariant. When an image $\mathcal{X}$ is given as input to the network, the associated input $I$ of the CPPN is obtained by extracting the images at the same coordinates in volumes $X_c, Y_c, Z_c$ anb by calculating their squares and component-wise product. Examples of U-net's inputs are given at different $z$ in Fig.\ref{fig:CPPN_in_all_CIBM}.b, the associated CPPN's inputs are shown in Fig.\ref{fig:CPPN_in_all_CIBM}.a. 
	
	\subsubsection{Evaluation of the CPPN influence}
	U-net and V-net label achieve pixel-wise classification based on contextual information and do not use location information. The distance at which a network can combined contextual information from a pixel is called the receptive field. To evaluate the benefit of the CPPN, we varied the receptive field of U-net and V-net by modifying their number of layers. The fewer the number of layer, the less contextual information the networks can use to achieve pixel-wise classification. In the case where networks have too little contextual information, they may confuse brain structures that look similar but are not located in the same place. If the CPPN learns the location of the CVS, this information would reduce the false-positives when networks have few layers. The number of layers of U-net and V-net were increased from 9 to 31 and from 7 to 24 in the case of U-net and V-net respectively.
	
	In the 2D case the number of layers was increased by adding, at the bottom of the U, a max pooling operation followed by two $3\times3$ convolutional layers, one $2\times2$ deconvolutional layer and two $3\times3$ convolutional layers (5 layers in total). In the 3D case we added, one $2\times2$ convolutional layer, three $3\times3$ convolutional layers, one $2\times2$ deconvolutional layer and finally three $3\times3$ convolutional layers (8 layers in total) were added at the bottom of the V. The original U-net (24 layers) and V-net (31 layers) are represented in Fig.\ref{fig:networks}.b and Fig.\ref{fig:networks}.a respectively. An increase in depth from 19 layers (Fig.\ref{fig:networks}.c) to 24 layers (Fig.\ref{fig:networks}.b) is given as an example for the U-net.
	
	\section{Experiments and results}
	
	\subsection{Training procedure pinciple}
	\subsubsection{Training procedure of U-net and V-net }
	During the training process, we gave batches composed of randomly drawn sub-volumes (coronal view) as inputs to the U-net and the V-net as well as the corresponding normalized sub-volumes’ coordinates to the CPPN. The targeted output was the corresponding label sub-volume. The optimization of the network’s parameters was performed following the cross-entropy loss during the first $5000$ steps to ensure a correct optimization start. It was then performed in accordance with the softDice loss ($\mathcal{L}$) \eqref{eqn2} in order to reach a better local minimum \cite{milletari2016v}.
	
	\begin{equation}
	\begin{array}{ccc}
	\mathcal{L} (Y, \hat{Y}) & = & 1 - 2\frac{\overset{N}{\underset{i=1}{\sum}} y_i \hat{y}_i}{\delta + \sum y_i + \hat{y}_i}
	\end{array}
	\label{eqn2}
	\end{equation}
	
	$y_i \in \{0,1\}$ and $\hat{y_i} \in [0,1]$ represent respectively ${Y}$’s and $\hat{Y}$’s i-th voxel’s values, with ${Y}$ and $\hat{Y}$ each containing N voxels. $\delta$ was set to $10^{-10}$ to avoid division by zero. Dice was calculated over the entire validation set every $1000$ iterations. The learning procedure was stopped as soon as it would not decrease over $5000$ iterations in a row. The optimized network with the highest Dice was eventually retained. All codes were implemented on Python with Tensorflow \cite{abadi2016tensorflow} and all networks were trained using NVIDIA Tesla V100 GPU.

	\subsubsection{Training procedure of \citet{moeskops2016automatic} network}
	The network was trained with batches of $50$ patches (randomly drawn in coronal plan) which contained at least half of patches labeled as CVS. The optimization of the network was done with cross-entropy. Early stopping was used in the same way as for U-net and V-net. Nevertheless, to limit the total optimization time, the performance on the validation set was obtained by calculating the average cross-entropy over a subset of voxels.
	
	\subsubsection{Parameters initialization and optimization-parameters setting}
	Before starting the optimization, the network’s weights were initialized in accordance with a uniform distribution with Xavier initializer \cite{glorot2010understanding} and the networks’ bias were initialized at $0.01$. The input sub-volumes’ size were $1\times128\times128$ in U-net cases and $64\times128\times128$ in the V-net cases. Concerning \citet{moeskops2016automatic} network, input patch of size $75 \times 75$, $51 \times 51$ and $25 \times 25$ were used in the case of the MRI dataset and input patch of size $127 \times 127$, $101 \times 101$ and $75 \times 75$ were used in the case of the US dataset. The optimization was performed using Adam optimizer \cite{kingma2014adam} with $\beta_1 = 0.9$, $\beta_2 = 0.999$ and $\epsilon = 10^{-8}$. The learning rates were set to: $10^{-4}$ for the $10000$ first iterations, $2 \times 10^{-5}$ for the $10000$ following iterations and $5 \times 10^{-6}$ until the end of training.
	
	\subsection{Test procedure}
	\subsubsection{Test volume segmentation}
	In the U-net cases, the test volumes were segmented image by image (size $320\times320$) in coronal orientation. In the V-net cases, subvolumes of size $64\times320\times320$ with an overlapping of $75\%$ were segmented by the network, the output probabilities to belong to the CVS were summed and the classes resurging in the highest values were attributed to each voxel. Finaly, in the \citet{moeskops2016automatic} network case, the test volumes were segmented vector by vector (size $320$) along coronal axis $z$.

	\subsubsection{Evaluation metrics}
	The proposed networks’ segmentation accuracy was evaluated using Dice \eqref{eqn:Dice} and MAD \eqref{eqn:MAD}:
	
	\begin{equation}
	\begin{array}{ccc}
	Dice & = & 2\frac{|Y \cap \hat{\hat{Y}}|}{|Y| + |\hat{\hat{Y}}|}
	\end{array}
	\label{eqn:Dice}
	\end{equation}
	
	\begin{equation}
	\begin{array}{ccc}
	MAD & = & \frac{1}{2} \left(\frac{\underset{x \in \partial \hat{\hat{Y}}}{\sum} d(x,\partial Y)}{|\partial Y|} + \frac{\underset{x \in \partial Y}{\sum} d(x,\partial \hat{\hat{Y}})}{|\partial \hat{\hat{Y}}|}\right)
	\end{array}
	\label{eqn:MAD}
	\end{equation}
	
	where $\partial Y$ and $\partial \hat{\hat{Y}}$ corresponds to $Y$'s and $\hat{\hat{Y}}$'s boarders respectively, $|Y|$ gives the number of ones in the set $Y$-set and $d$ is the euclidean distance defined as:
	
	\begin{equation}
	\begin{array}{ccc}
	d(x, \partial Y) & = & \underset{y \in \partial Y}{\min} \left\Vert x - y \right\Vert_2
	\end{array}
	\label{eqn5}
	\end{equation}
	
	To complete Dice and MAD, two other metrics closer to the clinic were used: the absolute volume difference ($\Delta V_a$) and the relative volume difference ($\Delta V_r$). They are defined by \eqref{eqn:Delta_Va} and \eqref{eqn:Delta_Vr} respectively. 
 	
	\begin{equation}
		\begin{array}{ccc}
			\Delta V_a & = & V_Y - V_{\hat{\hat{Y}}}
		\end{array}
		\label{eqn:Delta_Va}
	\end{equation}
	
	\begin{equation}
		\begin{array}{ccc}
			\Delta V_r & = & \frac{\Delta V_a}{V_Y}
		\end{array}
		\label{eqn:Delta_Vr}
	\end{equation}
	
	Where $V_Y$ and $V_{\hat{\hat{Y}}}$ are the volume of the reference segmentation and the volume of the automatic segmentation respectively.
	
	\subsection{Comparison of the networks}
	\subsubsection{t-test}
	Optimizing the network’s parameters means solving a non convex optimization problem whose solution is a local minimum. There are few chances to reach the same local minimum twice when the optimization procedure is run several times given the facts that the network’s parameters are randomly initialized and that a stochastic batch gradient descent is used for the optimization process. Hence, the accuracy obtained over the test set can be different from one optimization to another even if the training and validation sets are the same. To take this variability into account, the training procedure was always performed 5 times for each case. To highlight significant difference between two given architectures, the distribution resulting from the above-described 5-time optimization process was considered for Dice and MAD. A t-test was performed between two given distributions to compare them. The difference was considered significant when the p-value ($p$) was inferior to 0.05. For instance, the Dice obtained over the test set for the 5 optimizations of the 24-layer U-net were $0.808$, $0.809$, $0.811$, $0.8$ and $0.81$ whereas results showed $0.823$, $0.824$, $0.822$, $0.821$ and $0.822$ for the 31-layer V-net. The t-test comparing these two distributions resulted in $p=7.6 \times 10^{-5} < 0.05$. Hence there is a significant difference for Dice between these two architectures. For example, such results are given in Table.\ref{tab:CNN_comp_US} and Table.\ref{tab:quantitative_results_CPPN_all}.
	
	\subsubsection{Best validation}
	In a given configuration where the training procedure had been repeated for the same architecture, we considered that the best network was the one that had obtained the best Dice on the validation set. The final accuracy of this network was then evaluated on the test set. For example, in the case of the 24-layer U-net and the 31-layer V-net, the quantitative results given in Table.\ref{tab:networks_bestval_all} and Table.\ref{tab:Unet_vs_Vnet_bestval_norm_dil} and the qualitative results presented in Figure.\ref{fig:qualitative_results_hD} were all obtained from the best 24-layer U-net and the best 31-layer V-net.
	
	\subsection{Intraobserver variability}
	IOV was measured on five patients from the US test set: three patients with normal ventricles and two patients with dilated ventricles. The CVS of each of these patients was segmented twice. The agreement between each couple of segmentations was measured with respect to Dice, MAD, $\Delta V_a$ and $\Delta V_r$. The results obtained in the case of the dilated and the normal ventricles are given in Table.\ref{tab:Unet_vs_Vnet_bestval_norm_dil}. IOV was good in both cases but it was lower in the case of the dilated ventricles (according to all the metrics except $\Delta V_a$). A particularly good agreement between the segmentations of the dilated ventricles was found with a Dice of $0.898 \pm 0.008$ and a $\Delta V_r = 4.4 \%$ ($0.41 \pm 0.05$ cm$^3$).
	
	\subsection{Four-fold cross validation}
	To show that the distribution of the volumes between the training set, the validation set and the test set (Table.\ref{tab:dataset}) did not bias our results, a cross validation was performed. This experiment was realized with the four-fold cross validation dataset described in Table.\ref{tab:dataset_cross_val}. It was performed such that each of the 4 sets was, in turn, considered as the test set and the 3 remaining sets as the training set. In each configuration, the validation set remained unchanged and the U-net was optimized five times. The mean Dice and the mean MAD obtained for each fold are given in Table.\ref{tab:cross_val}.
	
	\begin{table}[!h]
		\caption{Mean Dice and Mean MAD values over cross validation test set.}
		\centering
		\begin{tabular}{ccc}
			\hline
			\multirow{2}{*}{Set} & Dice & MAD (mm) \\ 
			& (average) & (average) \\ 
			\hline 
			Fold 1 & 0.811 $\pm$ 0.004 & 0.49 $\pm$ 0.04\\ 
			Fold 2 & 0.807 $\pm$ 0.002 & 0.54 $\pm$ 0.03\\ 
			Fold 3 & 0.828 $\pm$ 0.003 & 0.53 $\pm$ 0.03\\ 
			Fold 4 & 0.797 $\pm$ 0.005 & 0.51 $\pm$ 0.05\\ 
			\hline 
		\end{tabular} 
		\label{tab:cross_val} 
	\end{table}

	The Dice and MAD varied from one test set to an other, but these variations remained reasonable. The Dice ranged from $0.797 \pm 0.005$ (fold 4) to $0.828 \pm 0.003$ (fold 3) and the MAD from $0.49 \pm 0.04$ mm (fold 1) to $0.54 \pm 0.03$ mm (fold 2).
	
	\subsection{Influence of the number of training volumes}
	
	\begin{figure}[!t]
		\centering
		\includegraphics[scale=.55]{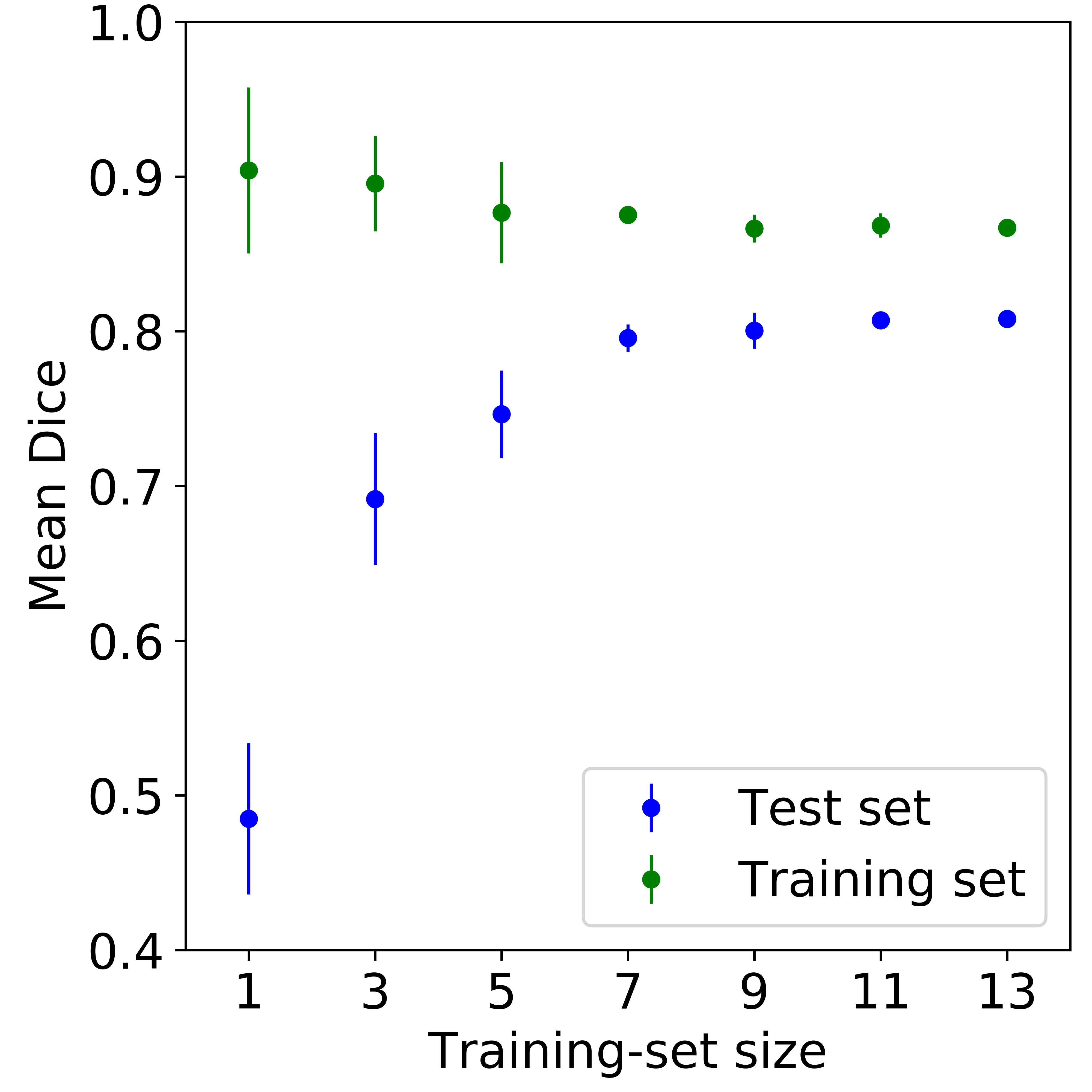}
		\caption{Mean Dice calculated over the training and the test sets as a function of the number of training volumes.}
		\label{fig:tss}
	\end{figure}
	
	Training a CNN for this application efficiently also requires understanding and determining the accurate number of images to be used in this process. That is why a 24 layers (Fig.\ref{fig:networks}.b) U-net with different training-set size was optimized. For this experiment, the dataset defined in Table.\ref{tab:dataset} was used, the validation and the test sets’ sizes remained the same while several sizes were used for the training sets: 1, 3, 5, 7, 9, 11 and 13. For each of these sizes, the optimization process was run 10 times with randomly-drawn volumes. The evolution of the mean Dice over the optimization as a function of the training set’s size is shown in Fig.\ref{fig:tss} for both the training and the test sets. Detailed values are given in Table.\ref{tab:tss_quantitative}.
	
	\begin{table}[!h]
		\caption{Mean Dice values over the training and test set values in relation to the training-set size.}
		\centering
		\begin{tabular}{cccc}
			\hline
			Training-set & Mean Dice & Mean Dice & \multirow{2}{*}{p-value} \\ 
			size & (training) & (test) & \\ 
			\hline
			1 & 0.904 $\pm$ 0.054 & 0.485 $\pm$ 0.049 & $1.22 \times 10^{-13}$\\ 
			3 & 0.895 $\pm$ 0.031 & 0.692 $\pm$ 0.043 & $1.97 \times 10^{-7}$\\ 
			5 & 0.877 $\pm$ 0.033 & 0.746 $\pm$ 0.028 & $4.81 \times 10^{-6}$\\ 
			7 & 0.875 $\pm$ 0.006 & 0.796 $\pm$ 0.009 & $2.08 \times 10^{-3}$\\ 
			9 & 0.866 $\pm$ 0.009 & 0.8 $\pm$ 0.012 & 0.091 \\ 
			11 & 0.868 $\pm$ 0.008 & 0.807 $\pm$ 0.004 & 0.685\\ 
			13 & 0.867 $\pm$ 0.006 & 0.808 $\pm$ 0.005 & 1\\ 
			\hline 
		\end{tabular} 
		\label{tab:tss_quantitative} 
	\end{table}	

	Results showed that mean Dices over the training and test sets decreased and increased respectively as the training set’s size increased. Both values seem to evolve slowly with the standard deviation being very low for training-set sizes reaching 7-or-more volumes. To determine whether the improvement was still significant, the results obtained for a training-set size of 13 were used as a baseline. Table.\ref{tab:tss_quantitative} shows that there is a significant difference for 1, 3, 5 and 7 patients but none for 9 ($p = 0.091 > 0.05$) and 11 patients ($p = 0.685 > 0.05$).
	
	\subsection{Comparison of the U-net, the V-net and \citet{moeskops2016automatic} network}
	 To evaluate the benefits of FCNs and to determine whether a 2D FCN or a 3D FCN architecture was more suitable for this problem, the segmentation time and the accuracy of the U-net, the V-net and \cite{moeskops2016automatic} were compared.\\
			
	\begin{table*}[!h] 
		\caption{Quantitative results given by the networks.} 
		\centering 
		\begin{tabular}{cccccc} 
			\hline 
			\multirow{2}{*}{Database} & \multirow{2}{*}{Network} & Dice & \multirow{2}{*}{p-value} & MAD (mm) & \multirow{2}{*}{p-value} \\ 
			& & (average) & & (average) & \\ \hline 
			\multirow{3}{*}{3D US} & V-net & \textbf{0.823 $\pm$ 0.001} & & \textbf{0.5 $\pm$ 0.03} & \\ 
			& U-net & 0.808 $\pm$ 0.004 & \textbf{7.59e-05} & 0.51 $\pm$ 0.02 & 0.7 \\ 
			& \citet{moeskops2016automatic} & 0.592 $\pm$ 0.019 & \textbf{8.96e-09} & 1.32 $\pm$ 0.1 & \textbf{3.36e-07} \\ 
			\\ 
			\multirow{3}{*}{3D MRI} & V-net & \textbf{0.892 $\pm$ 0.007} & & \textbf{0.51 $\pm$ 0.03} & \\ 
			& U-net & 0.889 $\pm$ 0.008 & 0.64 & 0.52 $\pm$ 0.04 & 0.82 \\ 
			& \citet{moeskops2016automatic} & 0.691 $\pm$ 0.027 & \textbf{5.34e-07} & 2.67 $\pm$ 0.42 & \textbf{6.79e-06} \\ 
			\hline 
		\end{tabular} 
		\label{tab:CNN_comp_US} 
	\end{table*}

	\begin{table*}[!h]
	\caption{Quantitative results obtained by the best networks over the whole US the test set compared to IOV}
	\centering
	\begin{tabular}{ccccc}
		\hline
		\multirow{2}{*}{Metrics} & \multicolumn{3}{c}{All CVS} \\ 
		& U-net & V-net & \citet{moeskops2016automatic} & IOV \\ 
		\hline 
		Dice & 0.81 $\pm$ 0.062 & \textbf{0.822 $\pm$ 0.053} & 0.621 $\pm$ 0.107 & 0.849 $\pm$ 0.041\\ 
		MAD (mm) & 0.5 $\pm$ 0.13 & \textbf{0.5 $\pm$ 0.13} & 1.16 $\pm$ 0.23 & 0.45 $\pm$ 0.07 \\ 
		$\Delta V_a$ (cm$^3$) & 0.38 $\pm$ 0.3 & \textbf{0.35 $\pm$ 0.28} & 3.3 $\pm$ 1.17 & 0.28 $\pm$ 0.14 \\ 
		$\Delta V_r$ (\%) & 11.9 $\pm$ 9.6 & \textbf{11.1 $\pm$ 10.1} & 101.5 $\pm$ 46.9 & 6.02 $\pm$ 3.09 \\ 
		\hline 
	\end{tabular} 
	\label{tab:networks_bestval_all} 
	\end{table*}

	\begin{table*}[!h]
		\caption{Quantitative results obtained by the best U-net and the best V-net over the normal and dilated ventricles of the test set compared to IOV}
		\centering
		\begin{tabular}{ccccccccccc}
			\hline
			\multirow{2}{*}{Metrics} & \multicolumn{5}{c}{Normal CVS} & \multicolumn{5}{c}{Dilated CVS} \\ 
			& U-net & V-net & & IOV & & & U-net & V-net & & IOV \\ 
			\hline 
			Dice & 0.776 $\pm$ 0.038 & \textbf{0.797 $\pm$ 0.041} & & 0.816 $\pm$ 0.009 & & & \textbf{0.893 $\pm$ 0.008} & 0.886 $\pm$ 0.004 & & 0.898 $\pm$ 0.008\\ 
			MAD (mm) & 0.55 $\pm$ 0.13 & \textbf{0.54 $\pm$ 0.13} & & 0.5 $\pm$ 0.04 & & & \textbf{0.38 $\pm$ 0.04} & 0.4 $\pm$ 0.06 & & 0.38 $\pm$ 0.04 \\ 
			$\Delta V_a$ (cm$^3$) & \textbf{0.35 $\pm$ 0.24} & \textbf{0.35 $\pm$ 0.29} & & 0.2 $\pm$ 0.11 & & & 0.45 $\pm$ 0.42 & \textbf{0.36 $\pm$ 0.24} & & 0.41 $\pm$ 0.05 \\ 
			$\Delta V_r$ (\%) & 14.5 $\pm$ 9.7 & \textbf{13.9 $\pm$ 10.6} & & 7.11 $\pm$ 3.6 & & & 5.4 $\pm$ 5.0 & \textbf{4.2 $\pm$ 3.1} & & 4.39 $\pm$ 0.02 \\ 
			\hline 
		\end{tabular} 
		\label{tab:Unet_vs_Vnet_bestval_norm_dil} 
	\end{table*}

	\subsubsection{Quantitative results}
	\paragraph{US dataset}
		According to the results presented in Table.\ref{tab:CNN_comp_US}, the V-net was significantly better than the U-net for Dice ($p = 7.6 \times 10^{-5}$) and also better fr MAD but not significantly ($p = 0.7$). It was also significantly better than the network proposed by \cite{moeskops2016automatic} for Dice ($p=8.96 \times 10^{-5}$) and MAD ($p=3.36 \times 10^{-7}$). Considering the accuracy of the best networks (Table.\ref{tab:networks_bestval_all}), the V-net was more accurate than the U-net for the Dice and as accurate for MAD. Both FCNs were extremely more accurate than \cite{moeskops2016automatic} according to all metrics . However, the comparison of the U-net and the V-net is more nuanced if we analyze the precision obtained on the dilated and non-dilated ventricles separately (Table.\ref{tab:Unet_vs_Vnet_bestval_norm_dil}). The V-net was more accurate than the U-net at segmenting the normal ventricles, but the U-net was slightly better than the V-net at segmenting dilated ventricles. According to the metrics that are closer to the clinic, the V-net is slightly more accurate or equivalent to the U-net. Considering all ventricles, a low $\Delta V_a$ was obtained by the U-net and the V-net ($0.35 \pm 0.28$ and $0.38 \pm 0.3$ respectively) with few differences between the normal and dilated ventricles. Consequently, $\Delta V_r$ was lower in the case of the dilated ventricles than for the normal ventricles, it was $13.9 \pm 10.6$ \% and $4.2 \pm 3.1$ \% in the case of the V-net.
		
	\paragraph{MRI dataset}
		As reported by the results presented in Table.\ref{tab:CNN_comp_US}, the V-net was more accurate than the U-net but not significantly. On the other hand, V-net was significantly better than the network proposed by \cite{moeskops2016automatic} according to Dice ($p=5.34 \times 10^{-7}$) and MAD ($p=6.79 \times 10^{-6}$). The Dice and the MAD obtained by the \cite{moeskops2016automatic} network were lower than the one reported in their article on the same database ($0.86 \pm 0.05$ and $0.52 \pm 0.25$ mm respectively). The accuracy obtained by applying the methodology described in \cite{moeskops2016automatic}: training with 7 classes (background, CVS, basal ganglia, white matter, brain stem, cortical grey matter, cerebellum) and application of a brain mask is reported in the Table.\ref{tab:CNN_comp_IRM_MonoMultiClass}. A similar value was then obtained for Dice ($0.844 \pm 0.053$), the value obtained for MAD remained more important ($0.89 \pm 0.22$ mm). This last point could be explained by the fact that the MAD formula used in our paper is different from the one used in \cite{moeskops2016automatic} but it was not mentioned by the authors.

	\subsubsection{Qualitative results}
\begin{figure*}[!h]
		\centering
		\includegraphics[scale=0.52]{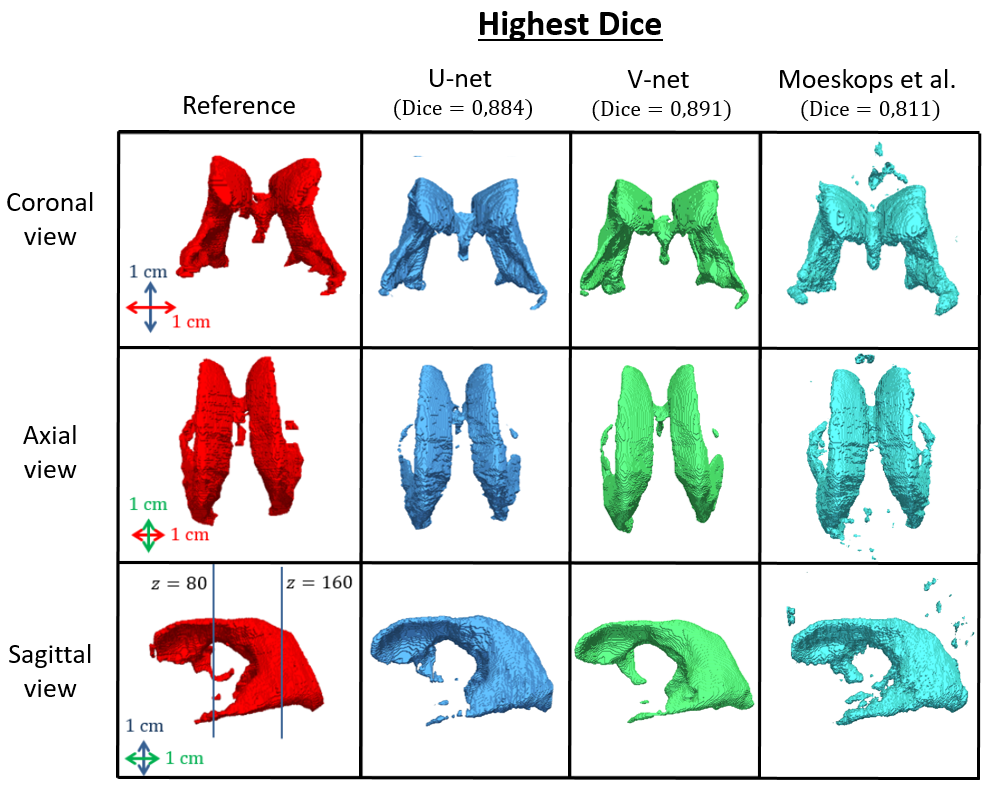}
		\caption{Qualitative segmentation results for the best networks on the test patient with the highest Dice.}
		\label{fig:qualitative_results_hD}
	\end{figure*}

\begin{figure*}[!h]
		\centering
		\includegraphics[scale=0.52]{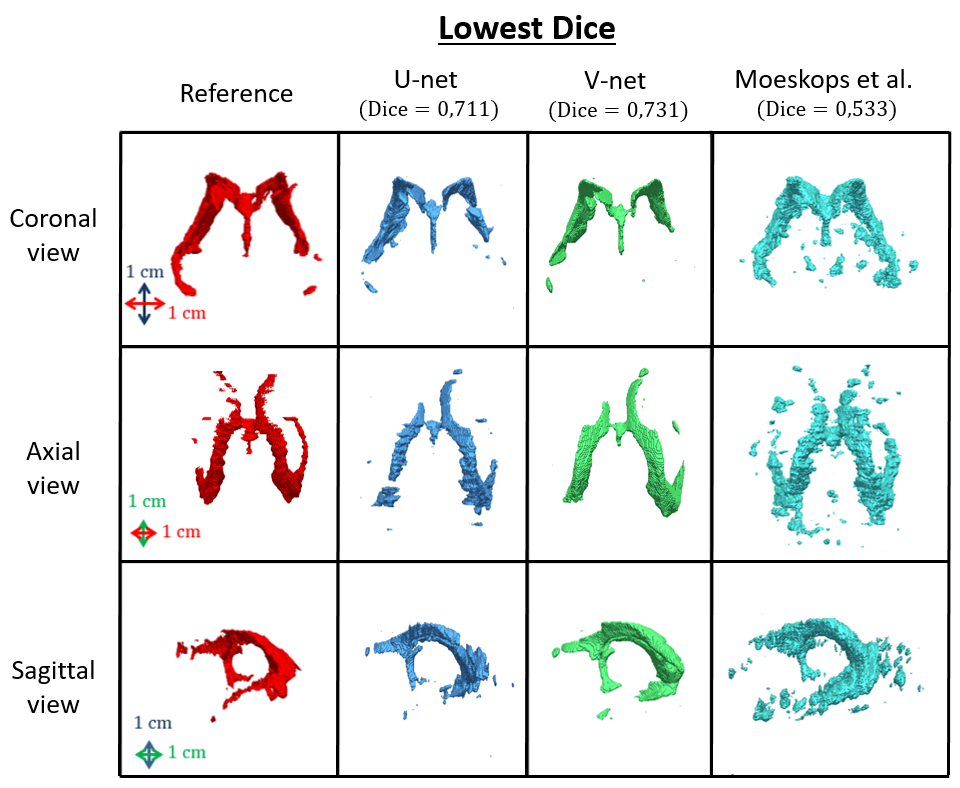}
		\caption{Qualitative segmentation results for the best networks on the test patient with the lowest Dice.}
		\label{fig:qualitative_results_lD}
	\end{figure*}

		Qualitative results for the networks with the highest Dice over the validation set can be observed for the test patients with the highest and lowest Dice (in the case of the V-net) respectively in Fig.\ref{fig:qualitative_results_hD} and Fig.\ref{fig:qualitative_results_lD}. The best test case, -a dilated CVS-, shows that 2D and 3D FCN architectures both performed well. However, the 3D FCN was better at segmenting the temporal horns, especially the left one. In addition, the 3D FCN architecture was slightly better at segmenting the third ventricle, but both architectures missed its thinnest part. The 2D non FCN has well identified the overall shape of the CVS, however there are more false positives than in the case of the FCN and the segmentation of the thinnest parts of the CVS are too wide. The worst test case, -a none dilated CVS-, shows that the 3D FCN architecture was better at segmenting the right temporal horn. However, both FCN architectures missed the left temporal horn. Overall, the segmentations showed similar results when it came to the CVS’s other parts except for the third ventricle that was better identified using the 3D architecture. Finally, the CVS segmented with the 3D FCN looked smoother and with fewer false positive particles. In the case of the 2D non FCN, all parts of the CVS were identified, in particular the right and left temporal horns. Nevertheless the segmentations produced were globally too wide and many false positives can be observed.

	\begin{table*}[!h] 
		\caption{Segmentation times of the networks depending on the GPU used.} 
		\centering 
		\begin{tabular}{cccc} 
			\hline 
			\multirow{3}{*}{Network} & \multicolumn{3}{c}{Segmentation time (s)} \\ 
			& Nvidia & Nvidia & Nvidia \\ 
			& Tesla V100 & 1080 TI & Quadro M1000M \\ 
			\hline 
			U-net & \textbf{3.5 $\pm$ 0.2} & 9.6 $\pm$ 0.2 & 70.4 $\pm$ 0.2 \\ 
			V-net & 10.2 $\pm$ 0.2 & 18.0 $\pm$ 0.0 & - \\ 
			\citet{moeskops2016automatic} & 7454.2 $\pm$ 101.6 & 10757.3 $\pm$ 28.8 & - \\ 
			\hline 
		\end{tabular} 
		\label{tab:Unet_vs_Vnet_seg_time} 
	\end{table*}
	
	\begin{table*}[!h] 
	\caption{Accuracy of the CVS segmentation for \citet{moeskops2016automatic} on the MRI dataset depending on training classes and brain mask use.}
	\centering 
	\begin{tabular}{cccc} 
		\hline 
		& CVS only & Multi-class & Multi-class + brain mask \\ \hline 
		Dice & 0.66 $\pm$ 0.19 & 0.801 $\pm$ 0.066 & \textbf{0.844 $\pm$ 0.053} \\ 
		MAD (mm) & 2.0 $\pm$ 0.93 & 1.65 $\pm$ 0.68 & \textbf{0.89 $\pm$ 0.22} \\ 
		\hline 
	\end{tabular} 
	\label{tab:CNN_comp_IRM_MonoMultiClass} 
	\end{table*}

	\subsubsection{Segmentation time}
	The segmentation times, given in Table.\ref{tab:Unet_vs_Vnet_seg_time}, were obtained using three GPUs: NVidia Tesla V100 (32 Go), Nvidia GTX 1080 (8 Go) and Nvidia Quadro M1000M (2 Go). Inferences were respectively performed with batches of size $80 \times 320 \times 320$, $20 \times 320 \times 320$ and $4 \times 320 \times 320$ in the case of the U-net. In the case of the V-net, inferences were always performed with batches of size $64 \times 320 \times 320$ and an overlapping of $75\%$ between batches. These results show that, whatever the GPU, the segmentations were performed faster by the U-net. Regarding \citet{moeskops2016automatic} network, inferences were performed with batches of size $1 \times 320$. The fastest segmentation time ($3.5 \pm 0.2$ s) was obtained using a Nvidia Tesla V100 GPU. V-net was not usable with the Nvidia Quadro M1000M because of insufficient memory ressources but achieved CVS segmentation in a few seconds with the other GPUs. No memory issues appended with the U-net, CVS segmentation was achieved in a time of $70.4 \pm 0.2$ s with the Quadro M1000M GPU which have very limited memory ressources. The slowest segmentation times were obtained by \cite{moeskops2016automatic}: more than two hours to segment one volume with the Nvidia Tesla V100 or the Nvidia 1080 TI. We did not perform segmentation with the Nvidia Quadro M1000M because segmentation time was too long. \\
	
	\subsection{The CPPN's influence}

	\begin{table*}[!h] 
		\caption{Influence of the CPPN on Dice and MAD for U-net and V-net depending on their number of layers.} 
		\centering 
		\begin{tabular}{ccccccccc} 
			\hline 
			\multirow{2}{*}{Database} & \multirow{2}{*}{Network} & \multirow{2}{*}{Layers} & \multicolumn{2}{c}{Dice (average)} & \multirow{2}{*}{p-value} & \multicolumn{2}{c}{MAD (mm) (average)} & \multirow{2}{*}{p-value} \\ 
		& 	& & no CPPN & CPPN & & no CPPN & CPPN & \\ 
			\hline 
			\multirow{9}{*}{3D US} & \multirow{4}{*}{U-net} & 24 & 0.808 $\pm$ 0.004 & 0.812 $\pm$ 0.003 & 0.15 & 0.51 $\pm$ 0.02 & 0.51 $\pm$ 0.01 & 0.87 \\ 
			& & 19 & 0.795 $\pm$ 0.005 & 0.807 $\pm$ 0.001 & \textbf{1.54e-03} & 0.55 $\pm$ 0.02 & 0.5 $\pm$ 0.02 & \textbf{2.91e-03} \\ 
			& & 14 & 0.762 $\pm$ 0.005 & 0.801 $\pm$ 0.002 & \textbf{3.17e-07} & 1.06 $\pm$ 0.08 & 0.54 $\pm$ 0.02 & \textbf{1.60e-06} \\ 
			& & 9 & 0.605 $\pm$ 0.008 & 0.757 $\pm$ 0.008 & \textbf{3.39e-09} & 3.62 $\pm$ 0.17 & 0.82 $\pm$ 0.06 & \textbf{1.42e-09} \\ 
			\\ 
			& \multirow{4}{*}{V-net} & 31 & 0.823 $\pm$ 0.001 & 0.824 $\pm$ 0.001 & 0.07 & 0.5 $\pm$ 0.03 & 0.52 $\pm$ 0.03 & 0.45 \\ 
			& & 23 & 0.821 $\pm$ 0.002 & 0.821 $\pm$ 0.002 & 0.94 & 0.55 $\pm$ 0.03 & 0.56 $\pm$ 0.04 & 0.53 \\ 
			& & 15 & 0.797 $\pm$ 0.001 & 0.808 $\pm$ 0.004 & \textbf{8.38e-04} & 0.85 $\pm$ 0.03 & 0.64 $\pm$ 0.02 & \textbf{1.06e-06} \\ 
			& & 7 & 0.518 $\pm$ 0.006 & 0.714 $\pm$ 0.006 & \textbf{4.94e-11} & 4.98 $\pm$ 0.17 & 1.31 $\pm$ 0.04 & \textbf{1.00e-10} \\ 
			\\ 
			\multirow{2}{*}{3D MRI} & \multirow{2}{*}{U-net} & 24 & 0.889 $\pm$ 0.008 & 0.9 $\pm$ 0.011 & 0.18 & 0.519 $\pm$ 0.038 & 0.48 $\pm$ 0.028 & 0.13 \\ 
			&         & 9 & 0.873$\pm$ 0.009 & 0.901 $\pm$ 0.013 & \textbf{6.76e-03} & 0.597 $\pm$ 0.048 & 0.464 $\pm$ 0.039 & \textbf{2.62e-03} \\
			\hline 
		\end{tabular} 
		\label{tab:quantitative_results_CPPN_all} 
	\end{table*}
	
	\begin{figure*}[h]
		\centering
		\includegraphics[scale=0.65]{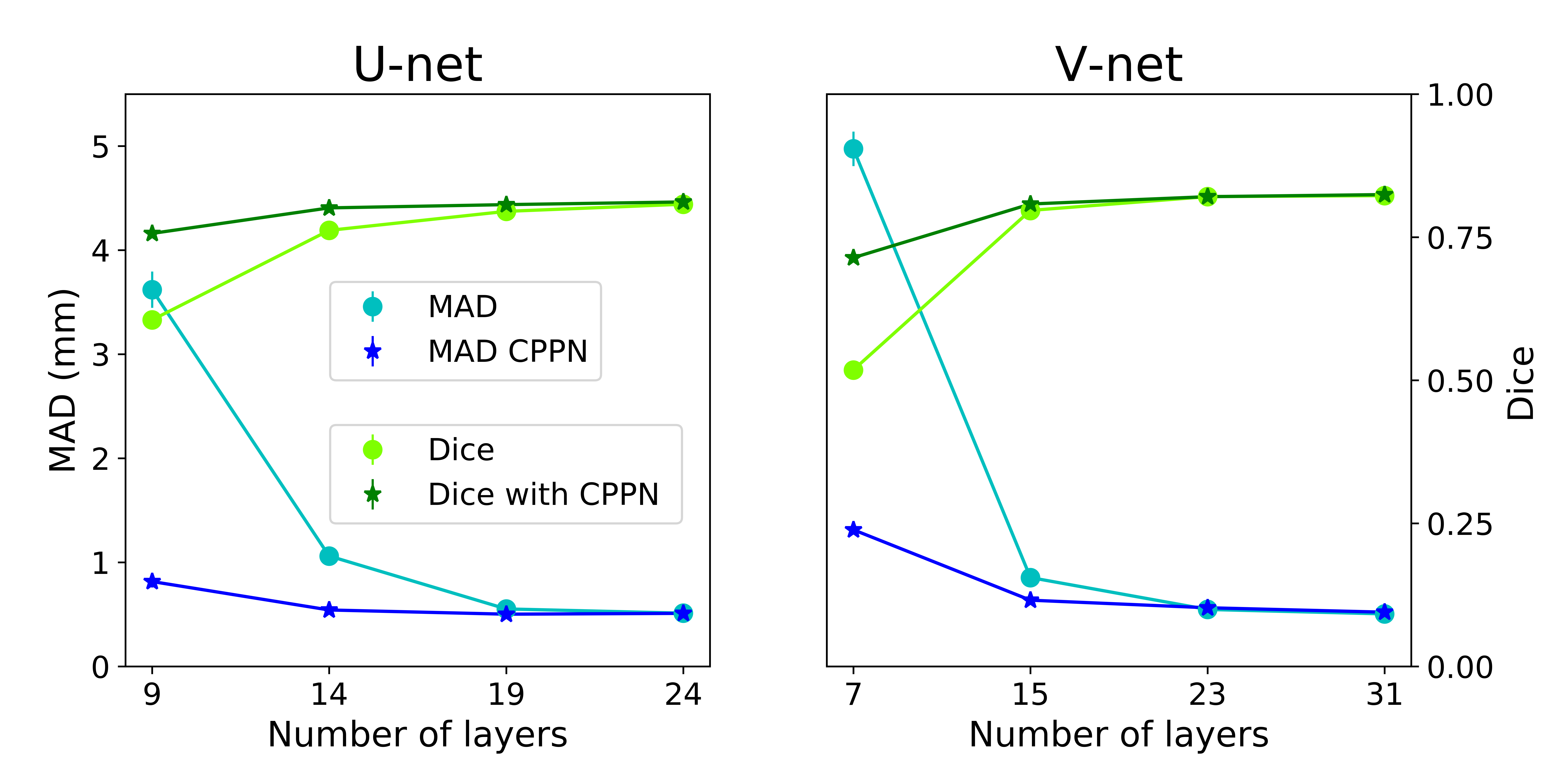}
		\caption{Mean results, for MAD and Dice, given by the 5 trainings of the U-net architectures.}
		\label{fig:MAD_Dice_CPPN}
	\end{figure*}
	
	\begin{figure*}[!h]
		\centering
		\includegraphics[scale=0.48]{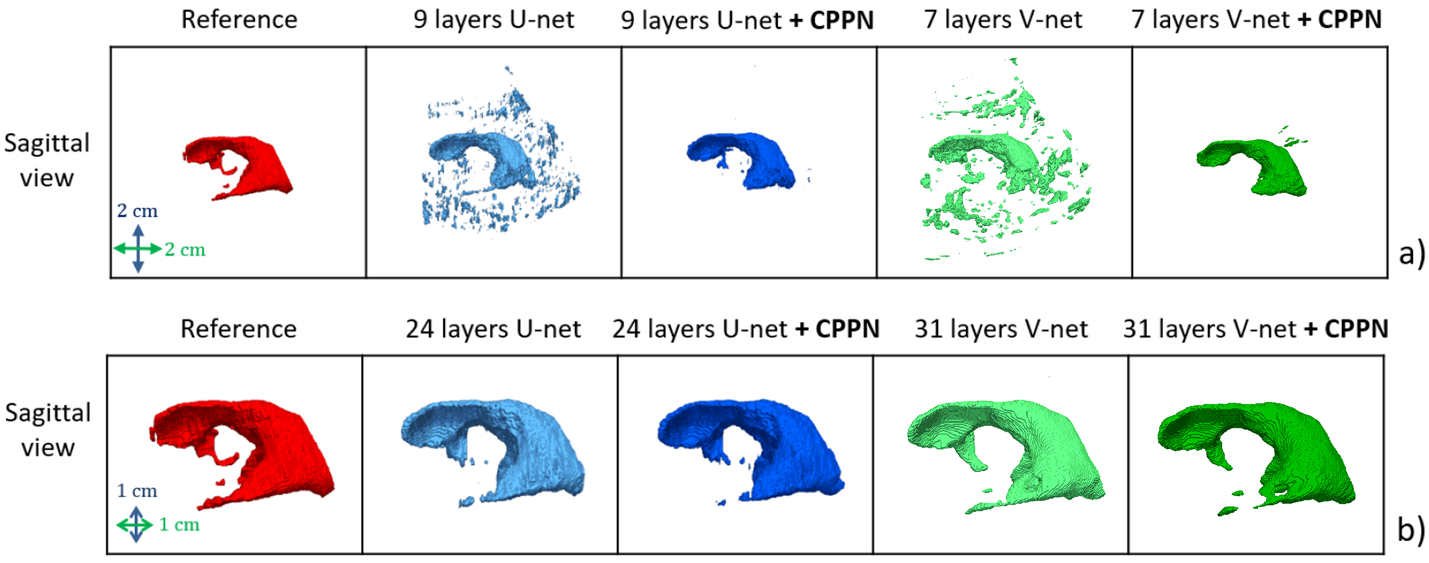}
		\caption{Qualitative results (best test patient) given by the 9-layer U-net and the 7-layer V-net a) and the 24-layer U-net and the 31-layer V-net b), with and	without CPPN.}
		\label{fig:qualitative_results_CPPN}
	\end{figure*}

	\begin{figure*}[!t]
		\centering
		\includegraphics[scale=0.5]{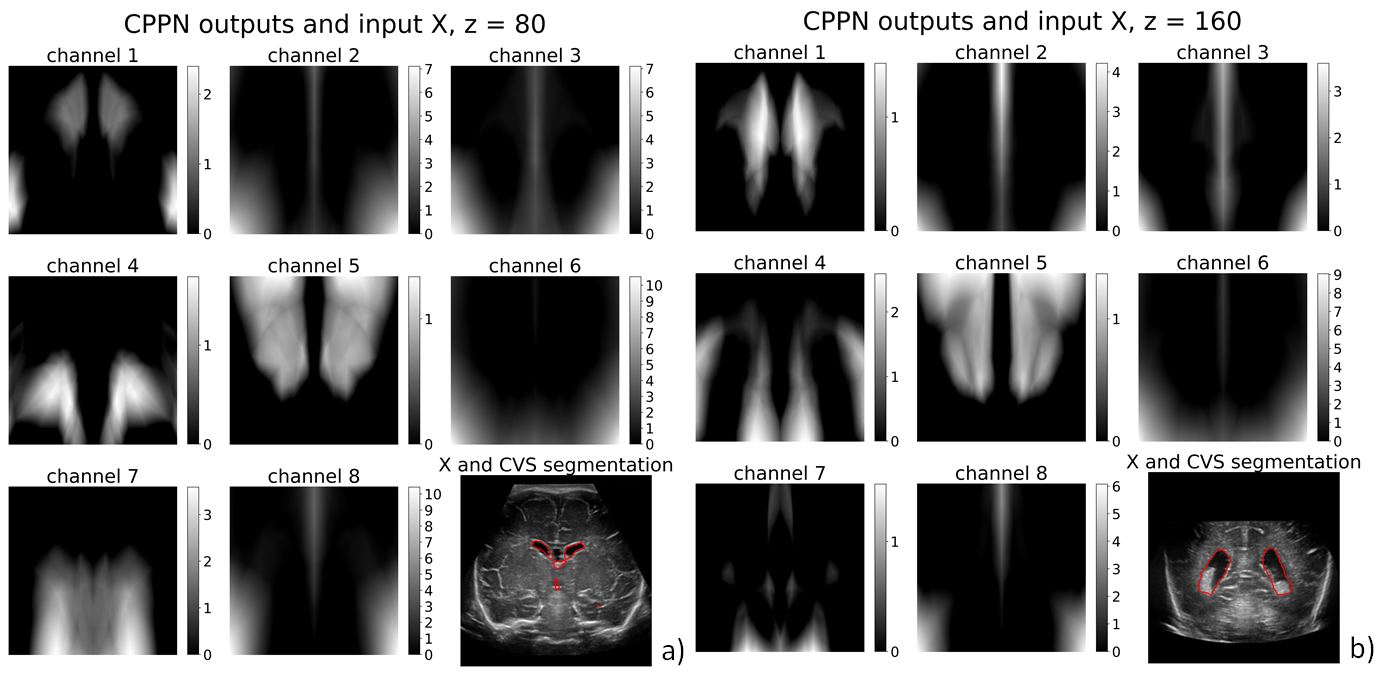}
		\caption{CPPN output with the corresponding labeled US image at a): $z = 80$ and b): $z = 160$.}
		\label{fig:CPPN_in_out}
	\end{figure*}	

	In this experiment, the number of layers of the U-net and the V-net were incrementally increased to determine if the CPPN brings information about the CVS location.\\
	
	This experiments’ quantitative results on the US dataset are given in Table.\ref{tab:quantitative_results_CPPN_all} and the evolution of MAD and Dice on US dataset as a function of the number of layers of U-net and V-net are represented in Fig.\ref{fig:MAD_Dice_CPPN}. A similar behaviour was observed in the 2D and 3D cases. As it can be seen in Fig.\ref{fig:MAD_Dice_CPPN}, with and without CPPN, Dice and MAD improved with the number of layers, although the improvment shrank as the number of layers increased. Nevertheless, it can be observed that the accuracy obtained with CPPN is better than without CPPN when the number of layers is low and equivalent when the number of layers is high. This result is confirmed by the quantitative values of Dice and MAD, and the p-values given in the Table.\ref{tab:quantitative_results_CPPN_all}. In the case of U-net, Dice and MAD were significantly ($p<0.05$) better with CPPN for the 9, 14 and 19-layers U-net. In the case of V-net, Dice and MAD were significantly better with CPPN for the 7-layers V-net and 15-layers V-net. The results also show that the networks without CPPN needed fewer layers to catch up with the performances of the networks with CPPN in the 3D case.\\

	The quantitative results obtained on the MRI dataset are in accordance with those obtained on the US dataset. Accuracy is given for a 24-layer U-net and a 9-layer U-net with and without CPPN in Table.\ref{tab:quantitative_results_CPPN_all}. Dice and MAD were better in all the cases where the CPPN was used and were significantly better in the case of the 9-layer U-net.\\

	Qualitative results for the best-performing 24-layer U-nets and the best-performing 31-layer V-nets associated with the patient from the US test set with the highest Dice (in the case of the 31-layer V-net) can be seen in Fig.\ref{fig:qualitative_results_CPPN}.b. These images show no significant differences between the segmentation performed with or without CPPN. Looking at the qualitative results for the best-performing 9-layer U-nets and the best-performing 7-layer V-nets associated with the patient from the test set with the highest Dice (Fig.\ref{fig:qualitative_results_CPPN}.a), a clear reduction of the false-positives is observed.

	\section{Discussion}
	
	\subsection{Four-fold cross validation}	
	
	The results given in Table.\ref{tab:cross_val} show that the maximum differences between two test sets for Dice and MAD are $0.031$ and $0.05$ mm respectively. These differences are small and the accuracy obtained in the worst cases is good, it is $0.797 \pm $0.005 for Dice and $0.54$ mm for MAD. This characterise the fact that each training set contained enough variability in order to enable the U-net to generalize its learning to an unknown data set. Based on these results, we conclude that the database used to compare the U-net and the V-net did not incorporate biases, such as very similar patients in the training set and the test set, which could have led to an overestimation of the results.
	
	\subsection{Influence of the training-set size}
	
	In this experiment, we studied the influence of the training-set size over the Dice values of the training set and the test set. Typically, these values respectively decrease and increase as the number training-set size increases. This is explained by the increasing variability contained in the training set leading to a fairly accurate generalization. These curves indicate which strategy should be used to improve the network’s accuracy over the test set: increasing the training-set size or designing a more suitable network architecture. In our case, test set’s Dice value did not significantly improve when increasing the training set’s size from 9 to 13 volumes, the training set’s Dice value with 13 volumes was slightly low ($0.867 \pm 0.006$) while the generalization gap for 13 volumes from the training set was slightly high ($0.059$). Based on these results, we conclude that the best strategy would be to design a more suitable architecture than the U-net for this problem. Increasing the training-set size would probably improve the network’s accuracy but it would be very time-consuming to gather a sufficient number of new volumes to obtain a significant improvement.
	
	\subsection{Comparison of the networks}
	In this part, the results obtained for accuracy and computing time are discussed in order to conclude on the use of the network of \cite{moeskops2016automatic}, the U-net and the V-net for this application. Possible explanations of the differences between the results given by these networks are also discussed.\\

	Compared according to the methodological metrics (Table.\ref{tab:CNN_comp_US}), the FCNs were significantly better than the network proposed by \cite{moeskops2016automatic}. They were therefore also better considering the clinical metrics, the differences obtained on volumes ($\Delta V_r = 101.5 \pm 46.9 \%$) make the use of this non-FCN network non suitable for clinical follow-up of CVS dilation. These poorer performances can be explained by the fact that, as it can be seen on Fig.\ref{fig:qualitative_results_hD} and Fig.\ref{fig:qualitative_results_lD}, the non-FCN network gave many false positives even though the pixels belonging to the CVS are very well identified. In particular the network seems to have difficulties to determine precisely the edges of the CVS. Similarly to what was shown on the MRI dataset (Table.\ref{tab:CNN_comp_IRM_MonoMultiClass}), training the networks with more classes and using a brain mask could reduce the number of false positives and improve the accuracy on the US dataset. Firstly, it could enable the network to learn to recognize other brain structures that may look similar to the CVS. Secondly, it could eliminate obvious false positives based on their location. However, these solutions are not suitable for 3D US because there are no manual annotations for structures other than CVS and the entire brain is not visible on the images.

	Considering the methodological metrics, the V-net was significantly better than the U-net for Dice and better for MAD. Moreover, the V-net has almost twice as few parameters as the U-net ($31 \times 10^6$ parameters for the U-net and $16 \times 10^6$ parameters for the V-net). This shows that the 3D V-net is able to encode information more efficiently than the 2D U-net and is overall more accurate for CVS segmentation. Nevertheless, according to the results given in the Table.\ref{tab:Unet_vs_Vnet_bestval_norm_dil} the gain in accuracy is due to the fact that the V-net was better than the U-net for segmenting normal ventricles, the U-net being more accurate for segmenting dilated ventricles. The qualitative results presented in Fig.\ref{fig:qualitative_results_lD} and Fig.\ref{fig:qualitative_results_hD} provide a possible explanation for this result. To achieve pixel-wise classification in our settings, the U-net used the information available in the coronal plan only whereas the V-net used the information available in all directions. This can results in 2D architectures struggling to segment some CVS's structures. In particular, the temporal horns, the frontal horns and the third ventricle are very thin in coronal slice when the ventricles are not dilated. Since a 3D architecture can analyze information in all space directions, it is easier for it to segment theses structures (Fig.\ref{fig:qualitative_results_lD}.b). In the case where ventricles are dilated, the structures are less thin in coronal slice and information available in this plan might be sufficient for a 2D architecture to perform as well as a 3D architecture (Fig.\ref{fig:qualitative_results_hD}.). Nevertheless, in both cases, the 3D kernels enabled to smooth CVS's edges contrary to 2D kernels, this is particularly striking in Fig.\ref{fig:qualitative_results_lD}.\\
	
	The difference between the IOV and the accuracy of the U-net and the V-net are given in the Table.\ref{tab:var_intra_op_vs_CNN}. By convention, the difference is negative when the network accuracy is better than the IOV and positive otherwise.
	
	\begin{table}[!h]
		\caption{Intraobserver variability in the segmentation of normal and dilated CVS.}
		\centering
		\begin{tabular}{ccccc}
			\hline
			\multirow{2}{*}{Metrics} & \multicolumn{2}{c}{Normal CVS} & \multicolumn{2}{c}{Dilated CVS} \\ 
			& U-net & V-net & U-net & V-net \\ 
			\hline 
			Dice & 0.04 & \textbf{0.019} & \textbf{0.005} & 0.011 \\ 
			MAD (mm) & 0.06 & \textbf{0.04} & \textbf{0} & 0.02 \\ 
			$\Delta V_a$ (cm$^3$) & \textbf{0.15} & \textbf{0.15} & 0.04 & \textbf{-0.05} \\ 
			\hline 
		\end{tabular} 
		\label{tab:var_intra_op_vs_CNN} 
	\end{table}

	The accuracy of the U-net and the V-net reached the IOV level, considering all metrics, in the case of the dilated ventricles. The accuracy of these segmentations is therefore excellent and it would be hard to improve in this case. In the case of normal ventricles, the V-net is globally closer to the IOV than the U-net with differences of $0.019$, $0.04$ mm and $0.15$ cm$^3$ considering Dice, MAD and $\Delta V_a$. The accuracy of these segmentations is therefore also excellent but it is still possible to improve the accuracy in this case. Performing data augmentation, such as horizontal symmetries, as proposed in \cite{martin2019segmentation} could be a way to bridge this gap.\\
	
	From a clinical point of view, the important question is whether these architectures enable the identification of a dilated ventricle and the monitoring of CVS dilation. The average volumes of normal and dilated ventricles in our database (Table.\ref{tab:dataset}) were respectively $2.7 \pm 0.8 $ cm$^3$ and $9.1 \pm 1.5 $ cm$^3$. The $\Delta V_a$ committed by the U-net and the V-net being less than $0.5$ cm$^3$ whether the ventricles are dilated or not, it can be concluded that these networks are capable of identifying a dilated ventricle unequivocally. Moreover, the segmentations of the dilated ventricles by the CNNs being at the level of a human observer, these networks would be able to follow the evolution of a ventricular dilation as precisely as a paediatrician performing manual segmentations.\\
		
	 The segmentation times (Table.\ref{tab:Unet_vs_Vnet_seg_time}) show that the FCN architectures are hundreds of times faster than the non-FCN architecture. This result is explained by the fact that FCNs are built to segment an entire image at each inference while the network proposed by \citet{moeskops2016automatic} can only segment one pixel. Therefore, it takes hundreds more inferences to non-FCN architectures to segment an entire image. Table.\ref{tab:Unet_vs_Vnet_seg_time} also reveals that the U-net was faster than the V-net at segmenting the CVS by a factor of $3$ with a Nvidia Tesla V100 GPU and by a factor of $2$ with a Nvidia 1080 TI GPU, while the V-net has fewer parameters than the U-net. This result can mainly be explained by the fact that the complete segmentation of a volume with a 3D network required an overlap between input batches to limit edge effects (along $z$ axis in our case). Thus, the segmentation of one volume required the U-net to segment 320 images and the V-net to segment 1088 images. Both architectures were able to segment the CVS in a clinical time of a few seconds using Tesla V100 GPU and 1080 TI GPU. Nevertheless, compared to the V-net, the U-net has the advantage to enable the realization of segmentations in a reasonable time ($70.4 \pm 0.2$ s) with hardware having limited memory resources (Nvidia Quadro M1000M) which are therefore less expensive.\\
	
	According to these results, FCN architectures should chosen over non-FCN architectures for this application. In addition, unless one wishes to segment normal ventricles accurately, the U-net is more suitable than the V-net for a clinical use. The U-net enables to follow ventricular dilation as accurately as a human observer performing manual segmentation. In addition, it enables automatic segmentations in a clinical time with GPUS which have few memory resources.
	D'après ses résultat
	\subsection{The influence of the CPPN}

\begin{figure}[!t]
		\centering
		\includegraphics[scale=0.47]{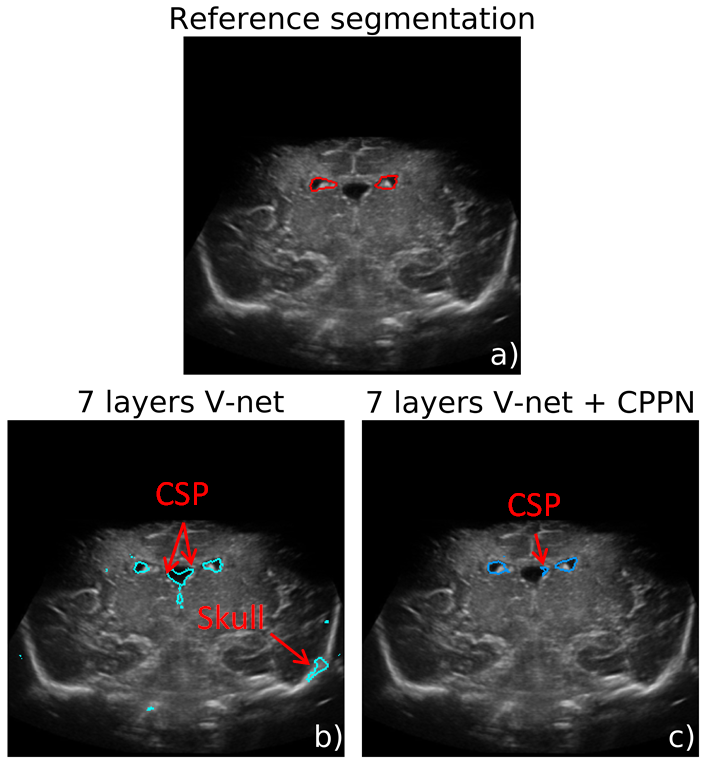}
		\caption{Reference segmentation a), 7 layers V-net automatic segmentation b) and 7 layers V-net + CPPN automatic segmentation c)}
		\label{fig:CPPN_qual_skull_csp}
	\end{figure}

	The CPPN was used to generate patterns (Fig.\ref{fig:CPPN_in_out}) which could provide location information to the networks. The quantitative results given in the Table.\ref{tab:quantitative_results_CPPN_all} show that the accuracy of U-net and V-net was better with CPPN when networks have few layers. This can be explained by the fact that the patterns generated by the CPPN enabled the networks to locate the characteristics in the common space defined for the database. Thus, when the networks were not able to use much contextual information, a decrease in the number of false positives, due to CVS-looking-like structures, was expected. This assumption was confirmed by Fig.\ref{fig:qualitative_results_CPPN}.a: many false positives were avoided when using the CPPN. Examples of false-positives from the best 7-layer V-net segmentation are given in Fig.\ref{fig:CPPN_qual_skull_csp}.b. It shows that the skull and the CSP, -looking like the plexus choroid and the CVS’s CSF respectively-, were segmented with no CPPN. Conversely, on adding the CPPN, the same V-net did not segment the skull at all and it segmented few of the CSP’s edges (Fig.\ref{fig:CPPN_qual_skull_csp}.c), the last of them being its closest part to the CVS geographically speaking.\\
	 
	The CPPN’s 8 output channel with the best 3D network, can be observed in Fig.\ref{fig:CPPN_in_out}. The patterns are shown for $z = 80$ (Fig.\ref{fig:CPPN_in_out}.a) and $z = 160$ (Fig.\ref{fig:CPPN_in_out}.b) with the network input $X$ (landmarks are given in Fig.\ref{fig:qualitative_results_hD}) and the CVS segmentation. Patterns on channels 1 and 5 show interesting results with shapes which may help to localize the CVS. When looking at the other shapes, they seem to partition the entire image into several parts. This may enable the network to use the patterns to know where the features are located and to avoid obvious false-positives.\\
	
	According to Table.\ref{tab:quantitative_results_CPPN_all} and Fig.\ref{fig:MAD_Dice_CPPN}, 2D networks might benefit more frome the CPPN than 3D networks. Indeed, the patterns generated by the CPPN are 3D so they can provide 3D location information. The 2D networks can only use the information contained in the coronal plans whereas 3D networks can utilize information in every direction. Thus, the CPPN is the only way for the 2D networks to use 3D information, which could explain why the 3D networks without CPPN needed fewer layers to catch up with the accuracy of the 3D networks with CPPN than the 2D networks.\\
	
	As a conclusion, the CPPN permits learning the CVS location and thus to decrease the number of false positives by providing this information to the networks. This is all the more obvious as the network’s number of layers decreases.
	
	\section{Conclusion}
	
	The first aim of this article was to measure the benefits of FCNs and to compare the potential of 2D and 3D CNNs to achieve CVS segmentation of preterm neonates in 3D US data. Our results showed that FCNs were extremely more accurate than non-FCNs on this task and that they could achieve segmentation hundreds of time faster. Our experiments also showed that a 3D FCN architecture is overall more accurate for this task. Nevertheless, a 2D FCN architecture is as accurate as a 3D FCN architecture for segmenting dilated ventricles, and in this case the segmentation accucary reaches the IOV. Moreover, a 2D FCN architecture enables to perform the segmentations in clinical time with hardwares that requires few memory resources and are therefore affordable. For all these reasons, a 2D FCN architecture may be preferable to a 3D FCN architecture in a clinical context where one wishes to monitor CVS dilation.
	
	In the future, these results will be compare to the accuracy obtained with MRI. Firstly, the match between the volumes measured in 3D US and MRI will be evaluated, as it was peformed by \cite{boucher2018computer}. Then, the intra and interobserver variability will be determined for these two modalities in order to establish the reliability of the measurements in both cases. Interobserver variability was measured in MRI by \cite{gui2012morphology} in the case of the segmentation of the CSF of two preterm neonates with dilated ventricles, its value ranged from $0.84$ to $0.89$. If we compare that result to the IOV, which is intrinsically higher than the interobserver variability, obtained over dilated ventricules in our case ($0.898 \pm 0.008$), it can be suggested that the measurements performed in 3D US images can be competitive with those performed in MRI.\\
	
	The second aim of this paper was to enable CNNs to encode CVS location since this information was used as an important features in most methods dedicated to the segmentation of preterm-neonates cerebral structures. Our experiments showed that the use of a CPPN enables the learning of a pattern dictionary encoding CVS location from the training dataset which improves the networks accuracy when they have few layers. Further investigations must be perform to evaluate the benefits of the CPPN for brain multi-structural segmentation. In addition, other CPPN inputs and basic mathematical functions could be tried to build a CPPN so as to better encode the location information into the patterns.\\
	
	To the best of our knowledge, this study is the first to compare the accuracy of 2D and 3D CNNs for the segmentation of preterm neonates' CVS in 3D US data and to distinguish between normal and dilated ventricles in the analysis of the results. This work is also the first to propose a modification of the U-net and the V-net architectures in order to learn a pattern dictionnary, for localisation, from training data. This study show that it is possible to accurately segment preterm neonates' CVS in a clinical time in 3D US images and that the accuracy can reach IOV in the case of dilated CVS. This work paves the way to the study of the clinical benefit of 3D US to monitor CVS dilation and the other cerebral structures of preterm neonates.
	
	\section*{Acknowledgments}
	This work was performed within the framework of the LABEX CELYA (ANR-10-LABX-0060) and PRIMES (ANR-11-LABX-0063) of Universite de Lyon, within the program "Investissements d'Avenir" (ANR-11-IDEX-0007) operated by the French National Research Agency (ANR). We would like to thank Dr. Borhane Slama, the innovation and clinical research commission president of CH Avignon, and the CH Avignon direction for their support to this research. 
	
	\bibliography{refs}

\begin{thebibliography}{46}
\providecommand{\natexlab}[1]{#1}
\providecommand{\url}[1]{\texttt{#1}}
\expandafter\ifx\csname urlstyle\endcsname\relax
  \providecommand{\doi}[1]{doi: #1}\else
  \providecommand{\doi}{doi: \begingroup \urlstyle{rm}\Url}\fi

\bibitem[Leviton and Gilles(1996)]{leviton1996ventriculomegaly}
Alan Leviton and Floyd Gilles.
\newblock Ventriculomegaly, delayed myelination, white matter hypoplasia, and
  “periventricular” leukomalacia: how are they related?
\newblock \emph{Pediatric neurology}, 15\penalty0 (2):\penalty0 127--136, 1996.

\bibitem[Burstein et~al.(1979)Burstein, Papile, and
  Burstein]{burstein1979intraventricular}
JEROME Burstein, LA~Papile, and ROCHELLE Burstein.
\newblock Intraventricular hemorrhage and hydrocephalus in premature newborns:
  a prospective study with ct.
\newblock \emph{American Journal of Roentgenology}, 132\penalty0 (4):\penalty0
  631--635, 1979.

\bibitem[Pappas et~al.(2018)Pappas, Adams-Chapman, Shankaran, McDonald, Stoll,
  Laptook, Carlo, Van~Meurs, Hintz, Carlson,
  et~al.]{pappas2018neurodevelopmental}
Athina Pappas, Ira Adams-Chapman, Seetha Shankaran, Scott~A McDonald, Barbara~J
  Stoll, Abbot~R Laptook, Waldemar~A Carlo, Krisa~P Van~Meurs, Susan~R Hintz,
  Martha~D Carlson, et~al.
\newblock Neurodevelopmental and behavioral outcomes in extremely premature
  neonates with ventriculomegaly in the absence of
  periventricular-intraventricular hemorrhage.
\newblock \emph{JAMA pediatrics}, 172\penalty0 (1):\penalty0 32--42, 2018.

\bibitem[Papile et~al.(1978)Papile, Burstein, Burstein, and
  Koffler]{papile1978incidence}
Lu-Ann Papile, Jerome Burstein, Rochelle Burstein, and Herbert Koffler.
\newblock Incidence and evolution of subependymal and intraventricular
  hemorrhage: a study of infants with birth weights less than 1,500 gm.
\newblock \emph{The Journal of pediatrics}, 92\penalty0 (4):\penalty0 529--534,
  1978.

\bibitem[Ancel et~al.(2006)Ancel, Livinec, Larroque, Marret, Arnaud, Pierrat,
  Dehan, Sylvie, Escande, Burguet, et~al.]{ancel2006cerebral}
Pierre-Yves Ancel, Florence Livinec, B{\'e}atrice Larroque, St{\'e}phane
  Marret, Catherine Arnaud, V{\'e}ronique Pierrat, Michel Dehan, N~Sylvie,
  Beno{\^\i}t Escande, Antoine Burguet, et~al.
\newblock Cerebral palsy among very preterm children in relation to gestational
  age and neonatal ultrasound abnormalities: the epipage cohort study.
\newblock \emph{Pediatrics}, 117\penalty0 (3):\penalty0 828--835, 2006.

\bibitem[Fox et~al.(2014)Fox, Choo, Rogerson, Spittle, Anderson, Doyle, and
  Cheong]{fox2014relationship}
Lisa~M Fox, Pauline Choo, Sheryle~R Rogerson, Alicia~J Spittle, Peter~J
  Anderson, Lex Doyle, and Jeanie~LY Cheong.
\newblock The relationship between ventricular size at 1 month and outcome at 2
  years in infants less than 30 weeks’ gestation.
\newblock \emph{Archives of Disease in Childhood-Fetal and Neonatal Edition},
  99\penalty0 (3):\penalty0 F209--F214, 2014.

\bibitem[Larroque et~al.(2003)Larroque, Marret, Ancel, Arnaud, Marpeau,
  Supernant, Pierrat, Roz{\'e}, Matis, Cambonie, et~al.]{larroque2003white}
B{\'e}atrice Larroque, S~Marret, Pierre-Yves Ancel, Catherine Arnaud, Loic
  Marpeau, Karine Supernant, V{\'e}ronique Pierrat, Jean-Christophe Roz{\'e},
  Jacqueline Matis, Gilles Cambonie, et~al.
\newblock White matter damage and intraventricular hemorrhage in very preterm
  infants: the epipage study.
\newblock \emph{The Journal of pediatrics}, 143\penalty0 (4):\penalty0
  477--483, 2003.

\bibitem[Melhem et~al.(2000)Melhem, Hoon~Jr, Ferrucci~Jr, Quinn, Reinhardt,
  Demetrides, Freeman, and Johnston]{melhem2000periventricular}
Elias~R Melhem, Alexander~H Hoon~Jr, Joseph~T Ferrucci~Jr, Cynthia~B Quinn,
  Elsie~M Reinhardt, Susan~W Demetrides, Bethany~M Freeman, and Michael~V
  Johnston.
\newblock Periventricular leukomalacia: relationship between lateral
  ventricular volume on brain mr images and severity of cognitive and motor
  impairment.
\newblock \emph{Radiology}, 214\penalty0 (1):\penalty0 199--204, 2000.

\bibitem[Maunu et~al.(2011)Maunu, Lehtonen, Lapinleimu, Matom{\"A}ki, Munck,
  Rikalainen, Parkkola, Haataja, and Group]{maunu2011ventricular}
Jonna Maunu, Liisa Lehtonen, Helena Lapinleimu, Jaakko Matom{\"A}ki, Petriina
  Munck, Hellevi Rikalainen, Riitta Parkkola, Leena Haataja, and PIPARI~Study
  Group.
\newblock Ventricular dilatation in relation to outcome at 2 years of age in
  very preterm infants: a prospective finnish cohort study.
\newblock \emph{Developmental Medicine \& Child Neurology}, 53\penalty0
  (1):\penalty0 48--54, 2011.

\bibitem[Mazzola et~al.(2014)Mazzola, Choudhri, Auguste, Limbrick, Rogido,
  Mitchell, and Flannery]{mazzola2014pediatric}
Catherine~A Mazzola, Asim~F Choudhri, Kurtis~I Auguste, David~D Limbrick, Marta
  Rogido, Laura Mitchell, and Ann~Marie Flannery.
\newblock Pediatric hydrocephalus: systematic literature review and
  evidence-based guidelines. part 2: management of posthemorrhagic
  hydrocephalus in premature infants.
\newblock \emph{Journal of Neurosurgery: Pediatrics}, 14\penalty0
  (Supplement\_1):\penalty0 8--23, 2014.

\bibitem[Leijser et~al.(2018)Leijser, Miller, van Wezel-Meijler, Brouwer,
  Traubici, van Haastert, Whyte, Groenendaal, Kulkarni, Han,
  et~al.]{leijser2018posthemorrhagic}
Lara~M Leijser, Steven~P Miller, Gerda van Wezel-Meijler, Annemieke~J Brouwer,
  Jeffrey Traubici, Ingrid~C van Haastert, Hilary~E Whyte, Floris Groenendaal,
  Abhaya~V Kulkarni, Kuo~S Han, et~al.
\newblock Posthemorrhagic ventricular dilatation in preterm infants: when best
  to intervene?
\newblock \emph{Neurology}, 90\penalty0 (8):\penalty0 e698--e706, 2018.

\bibitem[Brouwer et~al.(2012)Brouwer, de~Vries, Groenendaal, Koopman,
  Pistorius, Mulder, and Benders]{brouwer2012new}
Margaretha~J Brouwer, Linda~S de~Vries, Floris Groenendaal, Corine Koopman,
  Lourens~R Pistorius, Eduard~JH Mulder, and Manon~JNL Benders.
\newblock New reference values for the neonatal cerebral ventricles.
\newblock \emph{Radiology}, 262\penalty0 (1):\penalty0 224--233, 2012.

\bibitem[Davies et~al.(2000)Davies, Swaminathan, Chuang, and
  Betheras]{davies2000reference}
MW~Davies, M~Swaminathan, SL~Chuang, and FR~Betheras.
\newblock Reference ranges for the linear dimensions of the intracranial
  ventricles in preterm neonates.
\newblock \emph{Archives of Disease in Childhood-Fetal and Neonatal Edition},
  82\penalty0 (3):\penalty0 F218--F223, 2000.

\bibitem[Kishimoto et~al.(2016{\natexlab{a}})Kishimoto, de~Ribaupierre, Salehi,
  Romano, Lee, and Fenster]{kishimoto2016preterm}
Jessica Kishimoto, Sandrine de~Ribaupierre, Fateme Salehi, Walter Romano,
  David~SC Lee, and Aaron Fenster.
\newblock Preterm neonatal lateral ventricle volume from three-dimensional
  ultrasound is not strongly correlated to two-dimensional ultrasound
  measurements.
\newblock \emph{Journal of Medical Imaging}, 3\penalty0 (4):\penalty0 046003,
  2016{\natexlab{a}}.

\bibitem[Boucher et~al.(2018{\natexlab{a}})Boucher, Lipp{\'e}, Dupont, Knoth,
  Lopez, Shams, El-Jalbout, Damphousse, and Kadoury]{boucher2018computer}
Marc-Antoine Boucher, Sarah Lipp{\'e}, Caroline Dupont, Inga~Sophia Knoth,
  Gabriela Lopez, Roozbeh Shams, Ramy El-Jalbout, Am{\'e}lie Damphousse, and
  Samuel Kadoury.
\newblock Computer-aided lateral ventricular and brain volume measurements in
  3d ultrasound for assessing growth trajectories in newborns and neonates.
\newblock \emph{Physics in Medicine \& Biology}, 63\penalty0 (22):\penalty0
  225012, 2018{\natexlab{a}}.

\bibitem[Kishimoto et~al.(2016{\natexlab{b}})Kishimoto, Fenster, Salehi,
  Romano, Lee, and de~Ribaupierre]{kishimoto2016quantitative}
Jessica Kishimoto, Aaron Fenster, Fateme Salehi, Walter Romano, David~SC Lee,
  and Sandrine de~Ribaupierre.
\newblock Quantitative head ultrasound measurements to determine thresholds for
  preterm neonates requiring interventional therapies following
  intraventricular hemorrhage.
\newblock In \emph{Medical Imaging 2016: Ultrasonic Imaging and Tomography},
  volume 9790, page 97900S. International Society for Optics and Photonics,
  2016{\natexlab{b}}.

\bibitem[Moeskops et~al.(2016)Moeskops, Viergever, Mendrik, de~Vries, Benders,
  and I{\v{s}}gum]{moeskops2016automatic}
Pim Moeskops, Max~A Viergever, Adri{\"e}nne~M Mendrik, Linda~S de~Vries,
  Manon~JNL Benders, and Ivana I{\v{s}}gum.
\newblock Automatic segmentation of mr brain images with a convolutional neural
  network.
\newblock \emph{IEEE transactions on medical imaging}, 35\penalty0
  (5):\penalty0 1252--1261, 2016.

\bibitem[Liu et~al.(2017)Liu, Miller, Chau, and Studholme]{liu2017combining}
Mengyuan Liu, Steven~P Miller, Vann Chau, and Colin Studholme.
\newblock Combining spatial and non-spatial dictionary learning for automated
  labeling of intra-ventricular hemorrhage in neonatal brain mri.
\newblock In \emph{International Conference on Medical Image Computing and
  Computer-Assisted Intervention}, pages 789--797. Springer, 2017.

\bibitem[Qiu et~al.(2015{\natexlab{a}})Qiu, Yuan, Rajchl, Kishimoto, Chen,
  de~Ribaupierre, Chiu, and Fenster]{qiu20153d}
Wu~Qiu, Jing Yuan, Martin Rajchl, Jessica Kishimoto, Yimin Chen, Sandrine
  de~Ribaupierre, Bernard Chiu, and Aaron Fenster.
\newblock 3d mr ventricle segmentation in pre-term infants with
  post-hemorrhagic ventricle dilatation (phvd) using multi-phase geodesic
  level-sets.
\newblock \emph{NeuroImage}, 118:\penalty0 13--25, 2015{\natexlab{a}}.

\bibitem[Sciolla et~al.(2016)Sciolla, Martin, Quetin, and
  Delachartre]{sciolla2016segmentation}
Bruno Sciolla, Matthieu Martin, Philippe Quetin, and Philippe Delachartre.
\newblock Segmentation of the lateral ventricles in 3d ultrasound images of the
  brain in neonates.
\newblock In \emph{Ultrasonics Symposium (IUS), 2016 IEEE International}, pages
  1--4. IEEE, 2016.

\bibitem[Tabrizi et~al.(2018)Tabrizi, Obeid, Cerrolaza, Penn, Mansoor, and
  Linguraru]{tabrizi2018automatic}
Pooneh~R Tabrizi, Rawad Obeid, Juan~J Cerrolaza, Anna Penn, Awais Mansoor, and
  Marius~George Linguraru.
\newblock Automatic segmentation of neonatal ventricles from cranial ultrasound
  for prediction of intraventricular hemorrhage outcome.
\newblock In \emph{2018 40th Annual International Conference of the IEEE
  Engineering in Medicine and Biology Society (EMBC)}, pages 3136--3139. IEEE,
  2018.

\bibitem[Yasarla et~al.(2019)Yasarla, Wang, Hacihaliloglu, Patel,
  et~al.]{yasarla2019learning}
Rajeev Yasarla, Puyang Wang, Ilker Hacihaliloglu, Vishal~M Patel, et~al.
\newblock Learning to segment brain anatomy from 2d ultrasound with less data.
\newblock \emph{arXiv preprint arXiv:1912.08364}, 2019.

\bibitem[Qiu et~al.(2015{\natexlab{b}})Qiu, Yuan, Kishimoto, McLeod, Chen,
  de~Ribaupierre, and Fenster]{qiu2015user}
Wu~Qiu, Jing Yuan, Jessica Kishimoto, Jonathan McLeod, Yimin Chen, Sandrine
  de~Ribaupierre, and Aaron Fenster.
\newblock User-guided segmentation of preterm neonate ventricular system from
  3-d ultrasound images using convex optimization.
\newblock \emph{Ultrasound in medicine \& biology}, 41\penalty0 (2):\penalty0
  542--556, 2015{\natexlab{b}}.

\bibitem[Boucher et~al.(2018{\natexlab{b}})Boucher, Lippe, Damphousse,
  El-Jalbout, and Kadoury]{boucher2018dilatation}
Marc-Antoine Boucher, Sarah Lippe, Amelie Damphousse, Ramy El-Jalbout, and
  Samuel Kadoury.
\newblock Dilatation of lateral ventricles with brain volumes in infants with
  3d transfontanelle us.
\newblock \emph{arXiv preprint arXiv:1806.02305}, 2018{\natexlab{b}}.

\bibitem[Qiu et~al.(2017)Qiu, Chen, Kishimoto, de~Ribaupierre, Chiu, Fenster,
  and Yuan]{qiu2017automatic}
Wu~Qiu, Yimin Chen, Jessica Kishimoto, Sandrine de~Ribaupierre, Bernard Chiu,
  Aaron Fenster, and Jing Yuan.
\newblock Automatic segmentation approach to extracting neonatal cerebral
  ventricles from 3d ultrasound images.
\newblock \emph{Medical image analysis}, 35:\penalty0 181--191, 2017.

\bibitem[Martin et~al.(2018)Martin, Sciolla, Sdika, Wang, Quetin, and
  Delachartre]{martin2018automatic}
Matthieu Martin, Bruno Sciolla, Michaël Sdika, Xiaoyu Wang, Philippe Quetin,
  and Philippe Delachartre.
\newblock Automatic segmentation of the cerebral ventricle in neonates using
  deep learning with 3d reconstructed freehand ultrasound imaging.
\newblock In \emph{Ultrasonics Symposium (IUS), 2018 IEEE International}. IEEE,
  2018.

\bibitem[Badrinarayanan et~al.(2015)Badrinarayanan, Kendall, and
  Cipolla]{badrinarayanan2015segnet}
Vijay Badrinarayanan, Alex Kendall, and Roberto Cipolla.
\newblock Segnet: A deep convolutional encoder-decoder architecture for image
  segmentation.
\newblock \emph{arXiv preprint arXiv:1511.00561}, 2015.

\bibitem[Long et~al.(2015)Long, Shelhamer, and Darrell]{long2015fully}
Jonathan Long, Evan Shelhamer, and Trevor Darrell.
\newblock Fully convolutional networks for semantic segmentation.
\newblock In \emph{Proceedings of the IEEE conference on computer vision and
  pattern recognition}, pages 3431--3440, 2015.

\bibitem[Milletari et~al.(2016)Milletari, Navab, and Ahmadi]{milletari2016v}
Fausto Milletari, Nassir Navab, and Seyed-Ahmad Ahmadi.
\newblock V-net: Fully convolutional neural networks for volumetric medical
  image segmentation.
\newblock In \emph{3D Vision (3DV), 2016 Fourth International Conference on},
  pages 565--571. IEEE, 2016.

\bibitem[Ronneberger et~al.(2015)Ronneberger, Fischer, and
  Brox]{ronneberger2015u}
Olaf Ronneberger, Philipp Fischer, and Thomas Brox.
\newblock U-net: Convolutional networks for biomedical image segmentation.
\newblock In \emph{International Conference on Medical image computing and
  computer-assisted intervention}, pages 234--241. Springer, 2015.

\bibitem[Srhoj-Egekher et~al.(2013)Srhoj-Egekher, Benders, Viergever, and
  I{\v{s}}gum]{srhoj2013automatic}
Vedran Srhoj-Egekher, Manon~JNL Benders, Max~A Viergever, and Ivana
  I{\v{s}}gum.
\newblock Automatic neonatal brain tissue segmentation with mri.
\newblock In \emph{Medical Imaging 2013: Image Processing}, volume 8669, page
  86691K. International Society for Optics and Photonics, 2013.

\bibitem[Makropoulos et~al.(2014)Makropoulos, Gousias, Ledig, Aljabar, Serag,
  Hajnal, Edwards, Counsell, and Rueckert]{makropoulos2014automatic}
Antonios Makropoulos, Ioannis~S Gousias, Christian Ledig, Paul Aljabar, Ahmed
  Serag, Joseph~V Hajnal, A~David Edwards, Serena~J Counsell, and Daniel
  Rueckert.
\newblock Automatic whole brain mri segmentation of the developing neonatal
  brain.
\newblock \emph{IEEE transactions on medical imaging}, 33\penalty0
  (9):\penalty0 1818--1831, 2014.

\bibitem[Ghafoorian et~al.(2017)Ghafoorian, Karssemeijer, Heskes, van Uden,
  Sanchez, Litjens, de~Leeuw, van Ginneken, Marchiori, and
  Platel]{ghafoorian2017location}
Mohsen Ghafoorian, Nico Karssemeijer, Tom Heskes, Inge~WM van Uden, Clara~I
  Sanchez, Geert Litjens, Frank-Erik de~Leeuw, Bram van Ginneken, Elena
  Marchiori, and Bram Platel.
\newblock Location sensitive deep convolutional neural networks for
  segmentation of white matter hyperintensities.
\newblock \emph{Scientific Reports}, 7\penalty0 (1):\penalty0 1--12, 2017.

\bibitem[de~Brebisson and Montana(2015)]{de2015deep}
Alexander de~Brebisson and Giovanni Montana.
\newblock Deep neural networks for anatomical brain segmentation.
\newblock In \emph{Proceedings of the IEEE conference on computer vision and
  pattern recognition workshops}, pages 20--28, 2015.

\bibitem[Ganaye et~al.(2018)Ganaye, Sdika, and
  Benoit-Cattin]{ganaye2018towards}
Pierre-Antoine Ganaye, Micha{\"e}l Sdika, and Hugues Benoit-Cattin.
\newblock Towards integrating spatial localization in convolutional neural
  networks for brain image segmentation.
\newblock In \emph{2018 IEEE 15th International Symposium on Biomedical Imaging
  (ISBI 2018)}, pages 621--625. IEEE, 2018.

\bibitem[Stanley(2007)]{stanley2007compositional}
Kenneth~O Stanley.
\newblock Compositional pattern producing networks: A novel abstraction of
  development.
\newblock \emph{Genetic programming and evolvable machines}, 8\penalty0
  (2):\penalty0 131--162, 2007.

\bibitem[I{\v{s}}gum et~al.(2015)I{\v{s}}gum, Benders, Avants, Cardoso,
  Counsell, Gomez, Gui, H{\H{u}}ppi, Kersbergen, Makropoulos,
  et~al.]{ivsgum2015evaluation}
Ivana I{\v{s}}gum, Manon~JNL Benders, Brian Avants, M~Jorge Cardoso, Serena~J
  Counsell, Elda~Fischi Gomez, Laura Gui, Petra~S H{\H{u}}ppi, Karina~J
  Kersbergen, Antonios Makropoulos, et~al.
\newblock Evaluation of automatic neonatal brain segmentation algorithms: the
  neobrains12 challenge.
\newblock \emph{Medical image analysis}, 20\penalty0 (1):\penalty0 135--151,
  2015.

\bibitem[Marcus et~al.(2007)Marcus, Wang, Parker, Csernansky, Morris, and
  Buckner]{marcus2007open}
Daniel~S Marcus, Tracy~H Wang, Jamie Parker, John~G Csernansky, John~C Morris,
  and Randy~L Buckner.
\newblock Open access series of imaging studies (oasis): cross-sectional mri
  data in young, middle aged, nondemented, and demented older adults.
\newblock \emph{Journal of cognitive neuroscience}, 19\penalty0 (9):\penalty0
  1498--1507, 2007.

\bibitem[Landman and Warfield(2019)]{landman2019miccai}
Bennett~A Landman and Simon~K Warfield.
\newblock \emph{MICCAI 2012: Workshop on Multi-atlas Labeling}.
\newblock {\'e}diteur non identifi{\'e}, 2019.

\bibitem[Ioffe and Szegedy(2015)]{ioffe2015batch}
Sergey Ioffe and Christian Szegedy.
\newblock Batch normalization: Accelerating deep network training by reducing
  internal covariate shift.
\newblock \emph{arXiv preprint arXiv:1502.03167}, 2015.

\bibitem[Santurkar et~al.(2018)Santurkar, Tsipras, Ilyas, and
  Madry]{santurkar2018does}
Shibani Santurkar, Dimitris Tsipras, Andrew Ilyas, and Aleksander Madry.
\newblock How does batch normalization help optimization?
\newblock In \emph{Advances in Neural Information Processing Systems}, pages
  2483--2493, 2018.

\bibitem[Abadi et~al.(2016)Abadi, Barham, Chen, Chen, Davis, Dean, Devin,
  Ghemawat, Irving, Isard, et~al.]{abadi2016tensorflow}
Mart{\'\i}n Abadi, Paul Barham, Jianmin Chen, Zhifeng Chen, Andy Davis, Jeffrey
  Dean, Matthieu Devin, Sanjay Ghemawat, Geoffrey Irving, Michael Isard, et~al.
\newblock Tensorflow: A system for large-scale machine learning.
\newblock In \emph{12th $\{$USENIX$\}$ Symposium on Operating Systems Design
  and Implementation ($\{$OSDI$\}$ 16)}, pages 265--283, 2016.

\bibitem[Glorot and Bengio(2010)]{glorot2010understanding}
Xavier Glorot and Yoshua Bengio.
\newblock Understanding the difficulty of training deep feedforward neural
  networks.
\newblock In \emph{Proceedings of the thirteenth international conference on
  artificial intelligence and statistics}, pages 249--256, 2010.

\bibitem[Kingma and Ba(2014)]{kingma2014adam}
Diederik~P Kingma and Jimmy Ba.
\newblock Adam: A method for stochastic optimization.
\newblock \emph{arXiv preprint arXiv:1412.6980}, 2014.

\bibitem[Martin et~al.(2019)Martin, Sciolla, Sdika, Qu{\'e}tin, and
  Delachartre]{martin2019segmentation}
Matthieu Martin, Bruno Sciolla, Micha{\"e}l Sdika, Philippe Qu{\'e}tin, and
  Philippe Delachartre.
\newblock Segmentation of neonates cerebral ventricles with 2d cnn in 3d us
  data: suitable training-set size and data augmentation strategies.
\newblock In \emph{2019 IEEE International Ultrasonics Symposium (IUS)}, pages
  2122--2125. IEEE, 2019.

\bibitem[Gui et~al.(2012)Gui, Lisowski, Faundez, H{\"u}ppi, Lazeyras, and
  Kocher]{gui2012morphology}
Laura Gui, Radoslaw Lisowski, Tamara Faundez, Petra~S H{\"u}ppi, Fran{\c{c}}ois
  Lazeyras, and Michel Kocher.
\newblock Morphology-driven automatic segmentation of mr images of the neonatal
  brain.
\newblock \emph{Medical image analysis}, 16\penalty0 (8):\penalty0 1565--1579,
  2012.

\end{thebibliography}

\end{document}